\documentclass[aps,pra,10pt,twocolumn,showpacs, superscriptaddress]{revtex4-1} 
\usepackage{amsfonts,amsmath,amssymb}
\usepackage{graphicx}
\usepackage[pdfstartview=FitH,colorlinks=true,linkcolor=blue,citecolor=blue,urlcolor=blue]{hyperref}
\usepackage{tikz}
\usepackage{verbatim}

\begin{document}

\title{Thermodynamic behaviour of a one-dimensional Bose gas at low temperature}

\author{Giulia De Rosi}
\email{giulia.derosi@unitn.it}
\affiliation{INO-CNR BEC Center and Dipartimento di Fisica, Universit\`a di Trento, Via Sommarive 14, I-38123 Povo, Italy}

\author{Grigori E. Astrakharchik}
\email{grigori.astrakharchik@upc.edu}
\affiliation{Departament de F\'isica, Universitat Polit\`ecnica de Catalunya, 08034 Barcelona, Spain}

\author{Sandro Stringari}
\email{stringar@science.unitn.it}
\affiliation{INO-CNR BEC Center and Dipartimento di Fisica, Universit\`a di Trento, Via Sommarive 14, I-38123 Povo, Italy}

\date{\today}

\begin{abstract}
We show that the chemical potential of a one-dimensional (1D) interacting Bose gas exhibits a non-monotonic temperature dependence which is peculiar of superfluids. The effect is a direct consequence of the phononic nature of the excitation spectrum at large wavelengths exhibited by 1D Bose gases. 
For low temperatures $T$, we demonstrate that the coefficient in $T^2$ expansion of the chemical potential is entirely defined by the zero-temperature density dependence of the sound velocity. 
We calculate that coefficient along the crossover between the Bogoliubov weakly-interacting gas and the Tonks-Girardeau gas of impenetrable bosons. Analytic expansions are provided in the asymptotic regimes. 
The theoretical predictions along the crossover are confirmed by comparison with the exactly solvable Yang-Yang model in which the finite-temperature equation of state is obtained numerically by solving  Bethe-{\it ansatz} equations.
A 1D ring geometry is equivalent to imposing periodic boundary conditions and arising finite-size effects are studied in details.
At $T=0$ we calculated various thermodynamic functions, including the inelastic structure factor, as a function of the number of atoms, pointing out the occurrence of important deviations from the thermodynamic limit.
\end{abstract}

\pacs{PACS numbers}

\maketitle

\section{Introduction}
\label{Sec:Introduction}

It is well known that the thermodynamic behaviour of a superfluid is dominated, at low temperature, by the thermal excitation of phonons~\cite{Wilks1967}. This explains, in particular, the peculiar behaviour exhibited by the specific heat as well as by other fundamental thermodynamic functions. A non trivial (and less investigated in the literature) consequence of superfluidity shows up in the non-monotonic behaviour of the chemical potential~\cite{Papoular2012}. At low temperature $T$ the chemical potential increases with $T$ as a consequence of the thermal excitation of phonons. At high temperature, in the ideal gas classical regime, the chemical potential is instead a decreasing function of $T$. This non-monotonic behaviour has been recently measured in a strongly interacting atomic Fermi gas~\cite{Ku2012}, where it was shown that the chemical potential exhibits a maximum in the vicinity of the superfluid critical temperature.

It is consequently interesting to explore the low-temperature thermodynamic behaviour of other systems, like one-dimensional (1D) interacting Bose gases, which are known to exhibit a phononic excitation spectrum, despite the fact that they cannot be considered superfluids according to standard definition.
By investigating the drag flow caused by a moving external perturbation, Astrakharchik and Pitaevskii~\cite{Astrakharchik2004} have in fact shown that 1D Bose gases interacting with contact potential exhibit a traditional superfluid behaviour, characterized by the absence of friction force, only in the weakly interaction regime, where Bogoliubov theory applies and the gas can be locally considered Bose-Einstein condensed, despite the absence of true long range order.

In this work, we investigate the low-temperature expansion of the chemical potential $\mu$ of a 1D Bose gas with contact repulsive interaction for the whole crossover, ranging from the weakly to the strongly interaction limits. A major motivation is given by the possibility of comparing the low-T expansion of the chemical potential with the numerical results now available within the Yang-Yang theory \cite{Yang1969, Yang1970}, along the whole interaction strength crossover. Previous comparisons were in fact available only in the case of the Tonks-Girardeau limit \cite{Lang2015}, corresponding to the ideal Fermi gas, where the low-T expansion corresponds to the Sommerfeld expansion. We find that for all intermediate interaction regimes, described at $T=0$ by Lieb-Liniger (LL) theory, the increase of the chemical potential at low temperature follows the   $\mu \propto T^2$ law and is actually caused by the phononic nature of the long wavelength elementary excitations, as in usual superfluids~\cite{Papoular2012}. The relevant coefficient fixing the $T^2$ law depends on the density derivative of the $T=0$ sound velocity which is calculated using Lieb-Liniger theory. This feature strengthens the analogy with superfluids even in 1D dimension. Importantly, our results can be also generalized to every Luttinger liquid at low temperature whose macroscopic elementary excitations can be described in terms of phonons.

Recently, a ring geometry has been experimentally realized for a
microscopic system of $N=8 - 20$ atoms \cite{Labuhn2016}.
Motivated by the experimental progress, we study in details also the behavior of a gas containing a finite number of atoms in a ring, focusing on the deviations of its thermodynamic behavior from the one in the large $N$ limit.



Our system is a uniform gas of bosons interacting with a repulsive contact interaction
\begin{equation}
\label{Eq:Hamilt}
H = - \frac{\hbar^2}{2m}\sum_{i = 1}^N \frac{\partial^2}{\partial x_i^2} + 2c\sum_{i > j}^N\delta(x_i - x_j)
\end{equation}
where the interaction parameter $c$ is related to the 1D coupling constant $g_{\rm 1D} = -2\hbar^2/(m a_{\rm 1D})$ through $c = mg_{\rm 1D}/\hbar^2$, where $a_{\rm 1D}$ is the 1D scattering length.
The system~\eqref{Eq:Hamilt} has been realized experimentally for the whole interaction crossover by suitably tuning the interaction strength~\cite{Paredes2004,Kinoshita2006, Cazalilla2011}, described by the dimensionless parameter
\begin{equation}
\gamma = \frac{c}{n} = - \frac{2}{na_{\rm 1D}}
\label{Eq:gamma}
\end{equation}
from weak ($\gamma \rightarrow 0$) to strong ($\gamma \gg 1$) interactions~\cite{Kinoshita2006, Haller2009, Haller2010, Haller2011, Guarrera2012}. 
The Bogoliubov (BG) perturbative theory can be used in the limit of weak interactions.
In the Tonks-Girardeau (TG) limit of strong repulsions the bosons are impenetrable and their wave function can be mapped onto that of an ideal Fermi gas~\cite{Girardeau1960}.

The paper is organized as follows.

In Sec.~\ref{Sec:mu} we derive the low-temperature expansion of the chemical potential, starting from the free energy of an ideal phononic gas. This assumption is fully justified by the low-momenta behavior of the Lieb-Liniger excitation spectrum.
The low-temperature expansion exhibits a $T^2$-dependence on temperature, with the coefficient related to the density derivative of the LL sound velocity at zero temperature.
The Bethe-{\it ansatz} results for the chemical potential are shown to agree very well with the low-temperature expansion, for the whole BG-TG crossover.

In Sec.~\ref{Sec:BOG} we investigate the BG weakly-interacting gas. By considering the quantum fluctuation contribution in the ground-state energy at $T = 0$, we explore the behavior of the chemical potential and of the sound velocity. While this correction is important at $T = 0$, it does not affect the low-temperature expansion of the chemical potential.

Similarly to Sec.~\ref{Sec:BOG}, we calculate in Sec.~\ref{Sec:TG} the first corrections in the interaction parameter $\gamma$ to the TG strongly interacting gas. The starting point is the expansion, for large values of $\gamma$, of the ground-state energy of a hard-sphere gas.

In Sec.~\ref{Sec:compress} we derive the low-temperature expansions of both the adiabatic and the isothermal inverse compressibilities. The coefficients of the $T^2$ laws are studied as a function of the interaction parameter $\gamma$ and analytically calculated in the BG and TG limits.

In Sec.~\ref{Sec:pbc} we consider a ring configuration with a finite number of particles at zero temperature and calculate the finite-size corrections with respect to the thermodynamic limit for the energy, the chemical potential and the sound velocity. Results for the static inelastic structure factor for a finite number of particles are also reported.

In Sec.~\ref{Sec:Conclusion}, we draw our final conclusions.

\section{Low-temperature expansion of the chemical potential}
\label{Sec:mu}

\begin{figure}[ht]
\centering
\includegraphics[scale=0.5]{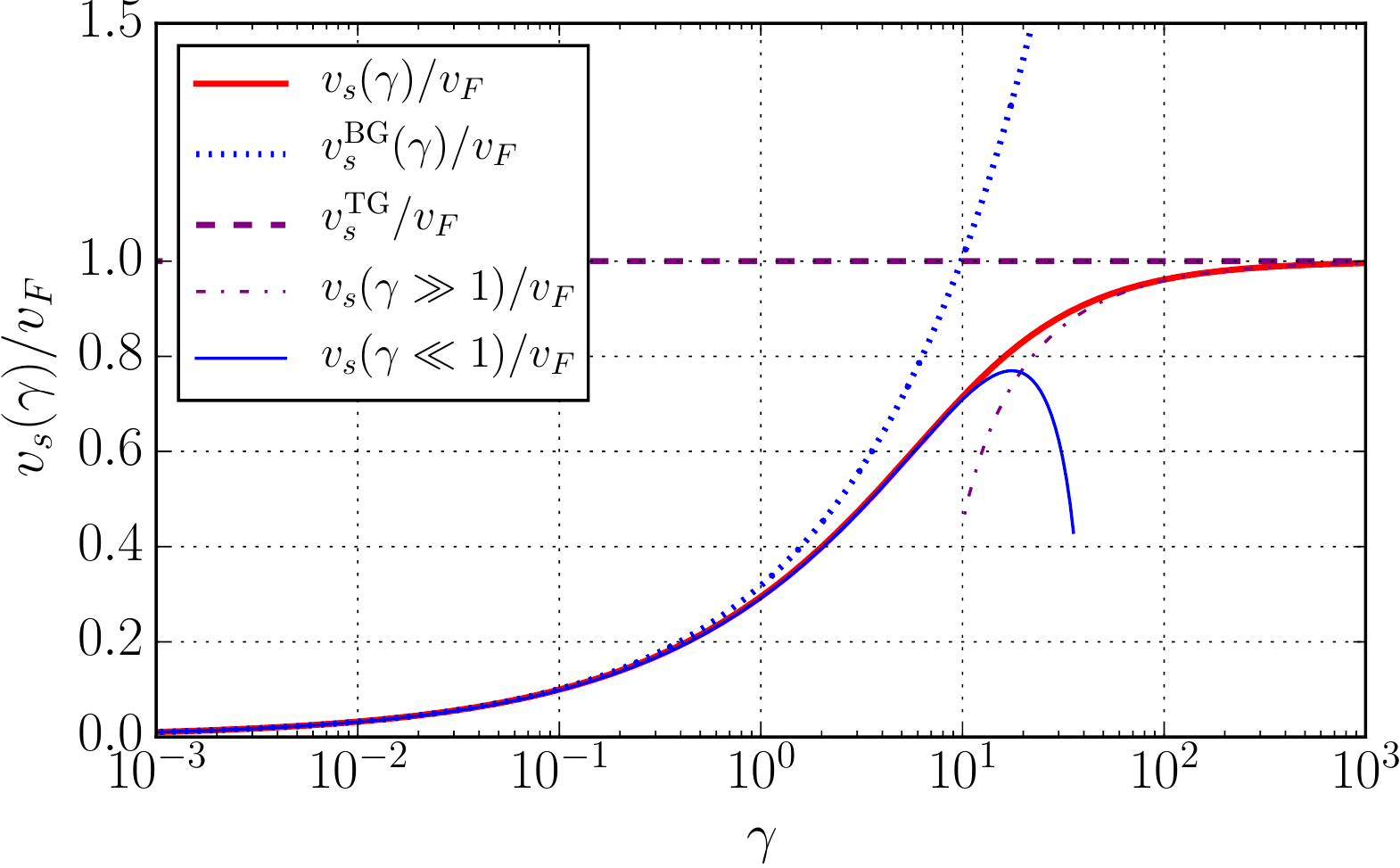}
\caption{(Color online) Sound velocity $v_s$ in units of Fermi velocity $v_F$ (solid line) as a function of the interaction parameter $\gamma$, calculated by solving the Lieb-Liniger equations. The Bogoliubov (dotted line, $v_s^{\rm BG}(\gamma)/v_F = \sqrt{\gamma}/\pi$) and Tonks-Girardeau (dashed line, $v_s^{\rm TG} = v_F$) limits, including their first-order corrections (thin solid line $v_s(\gamma \ll 1)/v_F = v_s^{\rm BG}(\gamma)/v_F \sqrt{1 - \sqrt{\gamma}/(2\pi)}$ and thin dot-dashed line $v_s(\gamma \gg 1)/v_F = \sqrt{1 - 8/\gamma}$, respectively) are present too, see Secs.~\ref{Sec:BOG} and~\ref{Sec:TG}. \label{fig:v}}
\end{figure}

It is well known that at $T=0$ the elementary excitations of an interacting 1D Bose gas have a phononic character at small momenta~\cite{Pitaevskii2016, Lieb1963}, characterized by the linear dispersion relation
\begin{equation}
\epsilon(p)_{p\to 0} = v_sp\;.
\label{Eq:vsp}
\end{equation}
At $T=0$ the sound velocity is related to the density dependence of the chemical potential according to the relation
\begin{equation}
v_s(\gamma)= \sqrt{\frac{n}{m}\frac{\partial\mu(T=0, \gamma)}{\partial n}}\;,
\label{Eq:vs}
\end{equation}
where $\mu$ is the chemical potential and $n=N/L$ denotes the linear density.
The density dependence of the chemical potential at zero temperature can be calculated within the Lieb--Liniger model.
The ratio between the sound velocity and the Fermi velocity $v_F=\pi \hbar n/m$ is known as the Luttinger parameter, $K_L = v_F/v_s$, and it plays an important role in defining the long-range properties of one-dimensional systems.
Figure~\ref{fig:v} shows the dependence of the sound velocity on the interaction parameter $\gamma$ for the Lieb-Liniger model, described by Hamiltonian~\eqref{Eq:Hamilt}.
There is a smooth crossover between the mean-field BG value defined as $mv_s^2 = g_{\rm 1D}n$ for weak interactions to the Tonks-Girardeau (ideal Fermi gas) value $v_s = v_F$ in the limit of strong repulsion.

\begin{figure}[ht]
\includegraphics[scale=0.442]{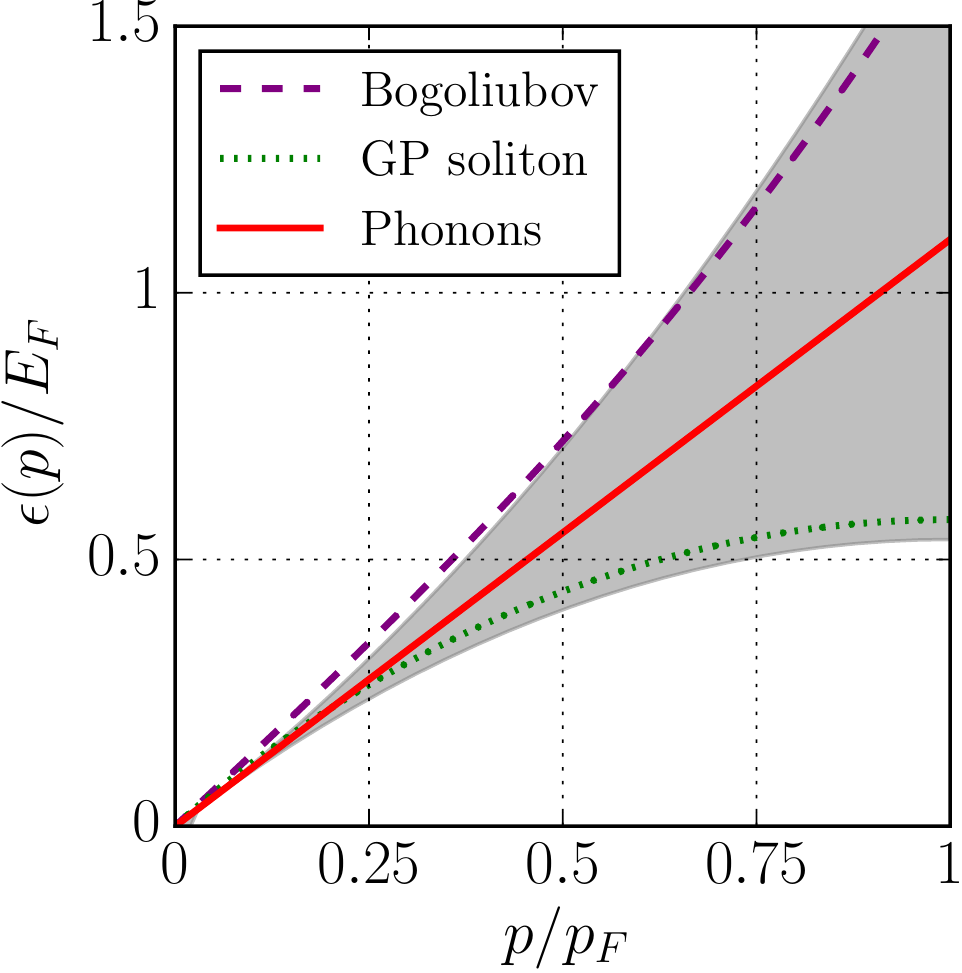}
\hspace{0.01cm}
\includegraphics[scale=0.442]{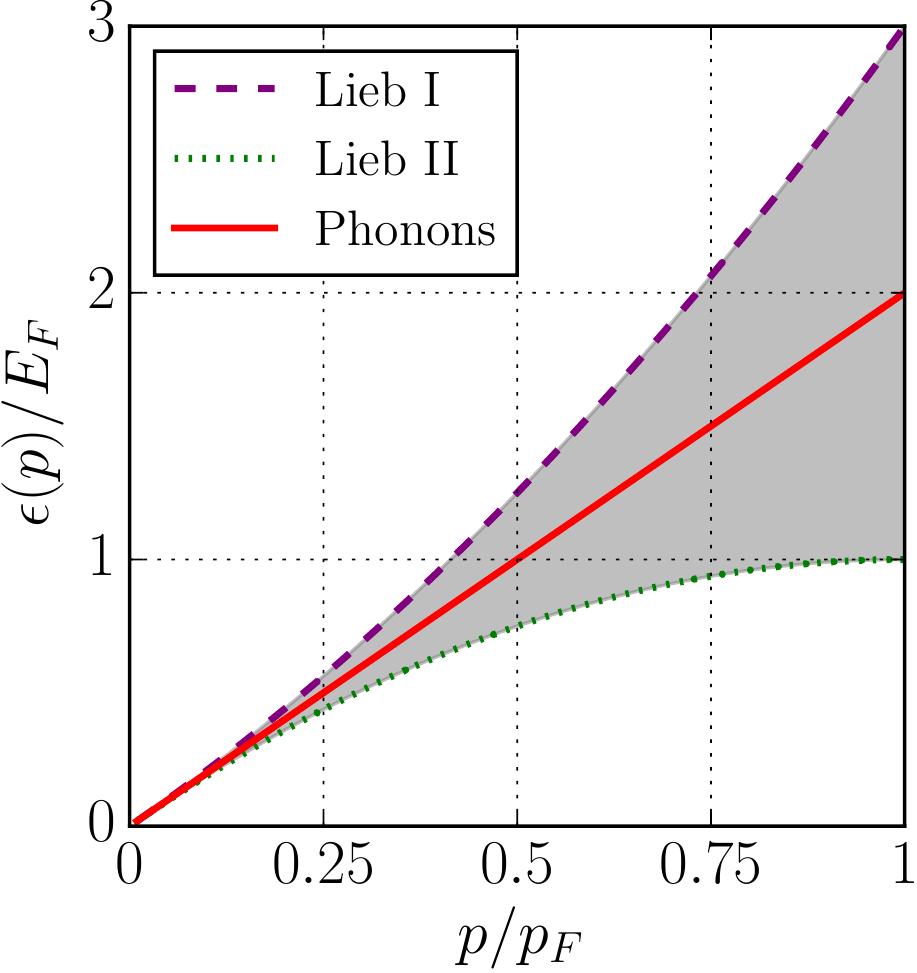}
\caption{(Color online) Lieb--Liniger excitation spectrum in the BG regime with $\gamma = 4.52$ (left) and in the deep TG regime with $\gamma =  \infty$ (right). The units are the Fermi energy $E_F$ and the Fermi momentum $p_F$. The shaded region represents the continuum of the excitations and is delimited by the upper (Lieb~I) and the lower (Lieb~II) branch of the spectrum. On the left, the Lieb~I and II branches are not reported. On the left, the dashed line gives the Bogoliubov dispersion and the dotted line gives the mean field soliton spectrum. In the limit $\gamma \to 0$, the Lieb~I branch tends to be equal to the Bogoliubov dispersion, while the Lieb II one coincides with the soliton spectrum. The solid line is the Lieb--Liniger phononic spectrum calculated with $\gamma = 4.52$.
On the right, Lieb~I and Lieb~II branches are reported and they coincide with the particle and hole ideal Fermi gas excitations, respectively. The solid line is the phononic spectrum calculated with the Fermi velocity. \label{fig:double}}
\end{figure}

For larger momenta the 1D excitation spectrum is characterized by a continuous structure, bounded by two branches of elementary excitations~\cite{Lieb1963, Yang1969, Pitaevskii2016}, which have been the object of recent measurements~\cite{Meinert2015, Fabbri2015}. For small values of $\gamma$, the Lieb-I particle-like branch corresponds to the Bogoliubov excitation spectrum~\cite{Lieb1963, Kulish1976, Pitaevskii2016}. The Lieb-II hole-like branch is instead associated in the weakly-interacting regime with the dark soliton dispersion predicted by Gross-Pitaevskii theory~\cite{Pitaevskii2016, Ishikawa1980, Kulish1976}.
The two branches merge into the phononic spectrum for $p \ll mv_s$, Fig.~\ref{fig:double}.

\begin{figure}[htbp]
\centering
\includegraphics[scale=0.55]{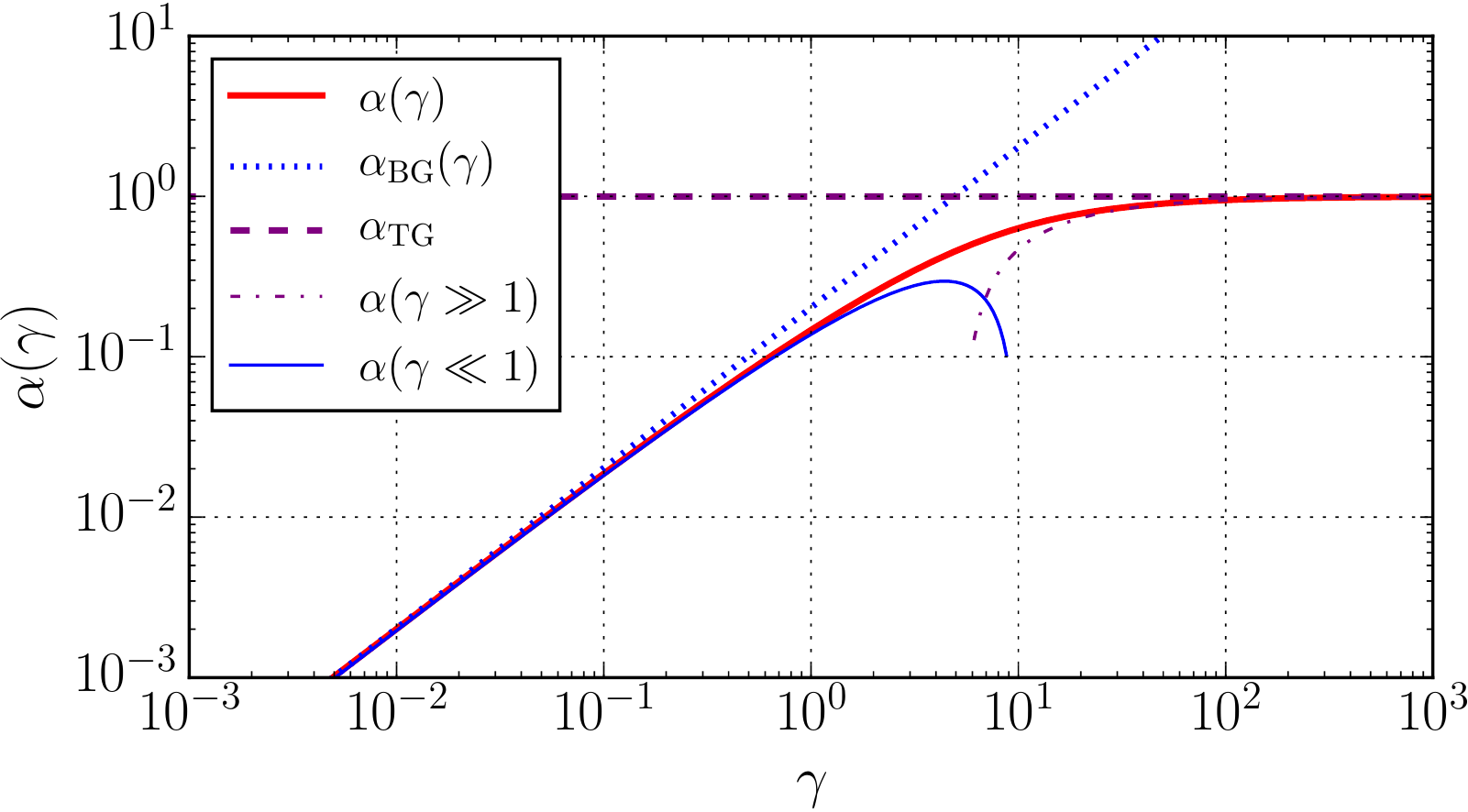}
\caption{(Color online) $\alpha(\gamma)$ (solid line) with the leading dependence (dashed line, $\alpha_{\rm TG} = 1$ and dotted line, $\alpha_{\rm BG}(\gamma) = 2\gamma/\pi^2$) and first order corrections [thin dot-dashed line, $\alpha(\gamma \gg 1) = 1 - 16/(3\gamma)$ and thin solid line, $\alpha(\gamma \ll 1) = \alpha_{\rm BG}(\gamma)(1 - \sqrt{\gamma}/\pi)$] for Tonks-Girardeau and Bogoliubov limits, respectively. \label{fig:alphacorr}}
\end{figure}

\begin{figure}[htbp]
\centering
\includegraphics[scale=0.55]{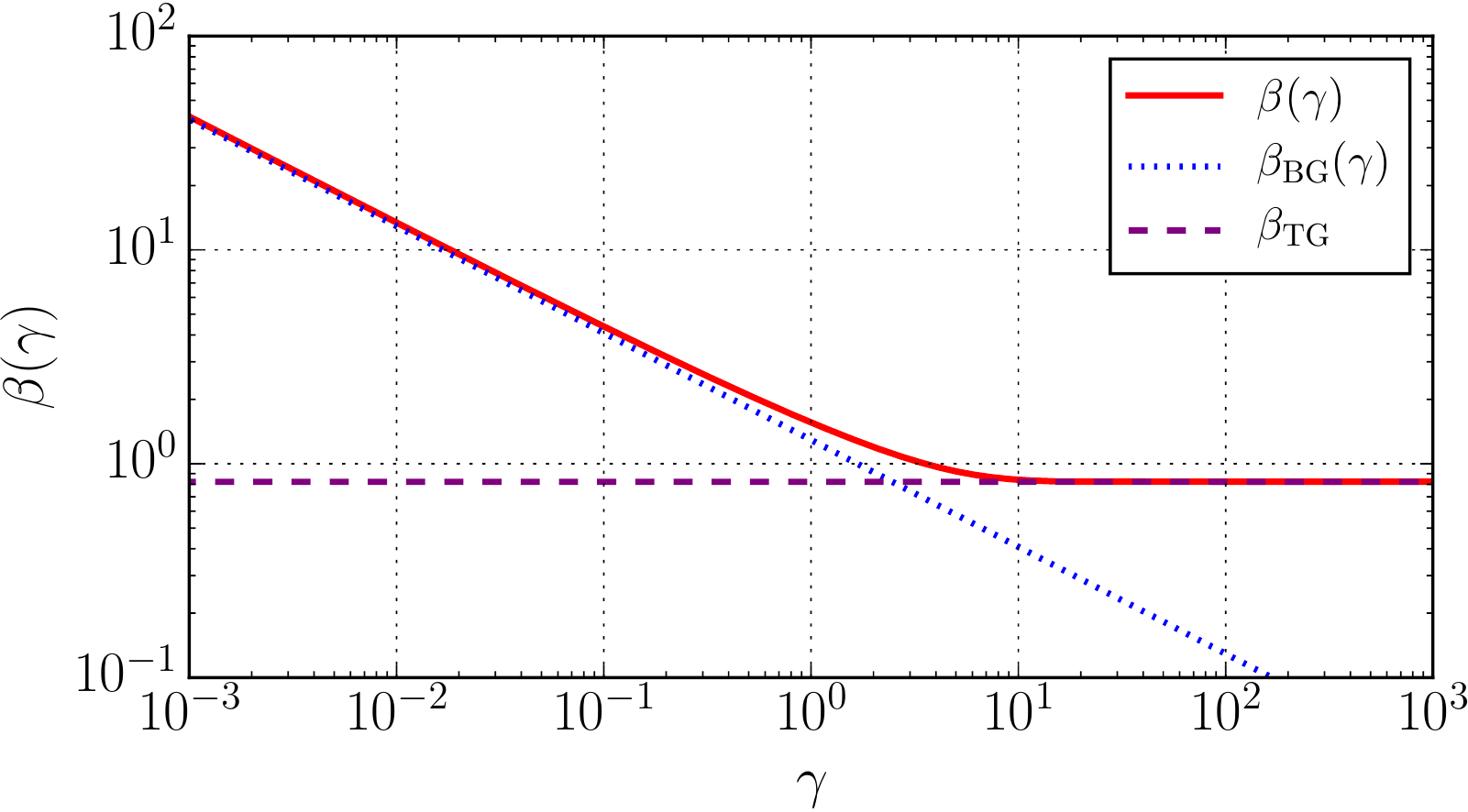}
\caption{(Color online) $\beta(\gamma)$ (solid line) with the Tonks-Girardeau (dashed line, $\beta_{\rm TG} = \pi^2/12$) and Bogoliubov (dotted line, $\beta_{\rm BG}(\gamma) = \pi^3/(24\sqrt{\gamma})$) limits. \label{fig:beta}}
\end{figure}

At low temperature ($k_BT \ll mv_s^2$) we expect that the thermodynamic behaviour of the system can be calculated in terms of a gas of non interacting phonons. The free energy $A = E - TS$ of this gas is then given by
\begin{equation}
\label{Eq:A}
A(T, L) = E_0 + \frac{k_BTL}{2\pi\hbar} \int_{-\infty}^{+\infty}\log \left[1 - e^{-\beta \epsilon(p)} \right]dp
\end{equation}
where $\epsilon(p)$ is dispersion~(\ref{Eq:vsp}) and we have added the energy $E_0$ calculated at $T = 0$ with the Lieb--Liniger theory. Notice that the thermal contribution to $A$ is affected by two-body interactions through the dependence of $\epsilon(p)$ on the interaction parameter $\gamma$. The integral of Eq.~(\ref{Eq:A}) yields the following low-temperature expansion for the free energy
\begin{equation}
\label{Eq:A_lowT}
A(T,L) = E_0 - \frac{\pi}{6} \frac{(k_BT)^2L}{\hbar v_s}\ ,
\end{equation}
which differs from the usual $T^4$ behaviour exhibited by three-dimensional ($3D$) superfluids~\cite{Pitaevskii2016} because of the 1D structure of the integral~(\ref{Eq:A}). Starting from result~(\ref{Eq:A_lowT}) for the free energy, one can calculate the low-temperature expansion of the chemical potential:
\begin{equation}
\label{Eq:mu general}
\mu(T, \gamma) = \left(\frac{\partial A}{\partial N}\right)_{T,L} = E_F\left[\alpha(\gamma) + \beta(\gamma)\left(\frac{T}{T_F}\right)^2 \right]
\end{equation}
where we have introduced the energy scale $E_F = k_BT_F = \hbar^2\pi^2n^2/(2m)$ given by the Fermi energy of a 1D Fermi gas, because it exhibits the same density dependence of the quantum degeneracy temperature of the system.
We have also defined the relevant dimensionless parameters of the expansion
\begin{equation}
\label{Eq:alpha}
\alpha(\gamma) = \frac{\mu(T = 0, \gamma)}{E_F}
\end{equation}
and
\begin{equation}
\label{Eq:beta}
\beta(\gamma) = \frac{\pi E_F}{6 \hbar v_s^2} \frac{\partial v_s}{\partial n} \ ,
\end{equation}
which are functions of the interaction parameter $\gamma$ and can be calculated at zero temperature using Lieb--Liniger theory. It is worth noticing that the parameter $\beta(\gamma)$, which is the most relevant because it fixes the leading coefficient of the low-$T$ expansion, depends on the density derivative of the sound velocity. The two numerical functions $\alpha(\gamma)$, Eq.~\eqref{Eq:alpha}, and $\beta(\gamma)$, Eq.~\eqref{Eq:beta}, have been calculated within LL theory and their values are reported in Figs.~\ref{fig:alphacorr} and \ref{fig:beta} with their BG and TG limits. In particular, the TG limits for $\alpha(\gamma)$ and $\beta(\gamma)$ reproduce the low-temperature Sommerfeld expansion of the chemical potential for the 1D ideal Fermi gas, Eq.~\eqref{Eq:Sommerfeld}. 

\begin{figure}[ht]
\centering
\includegraphics[scale=0.5, angle = 0]{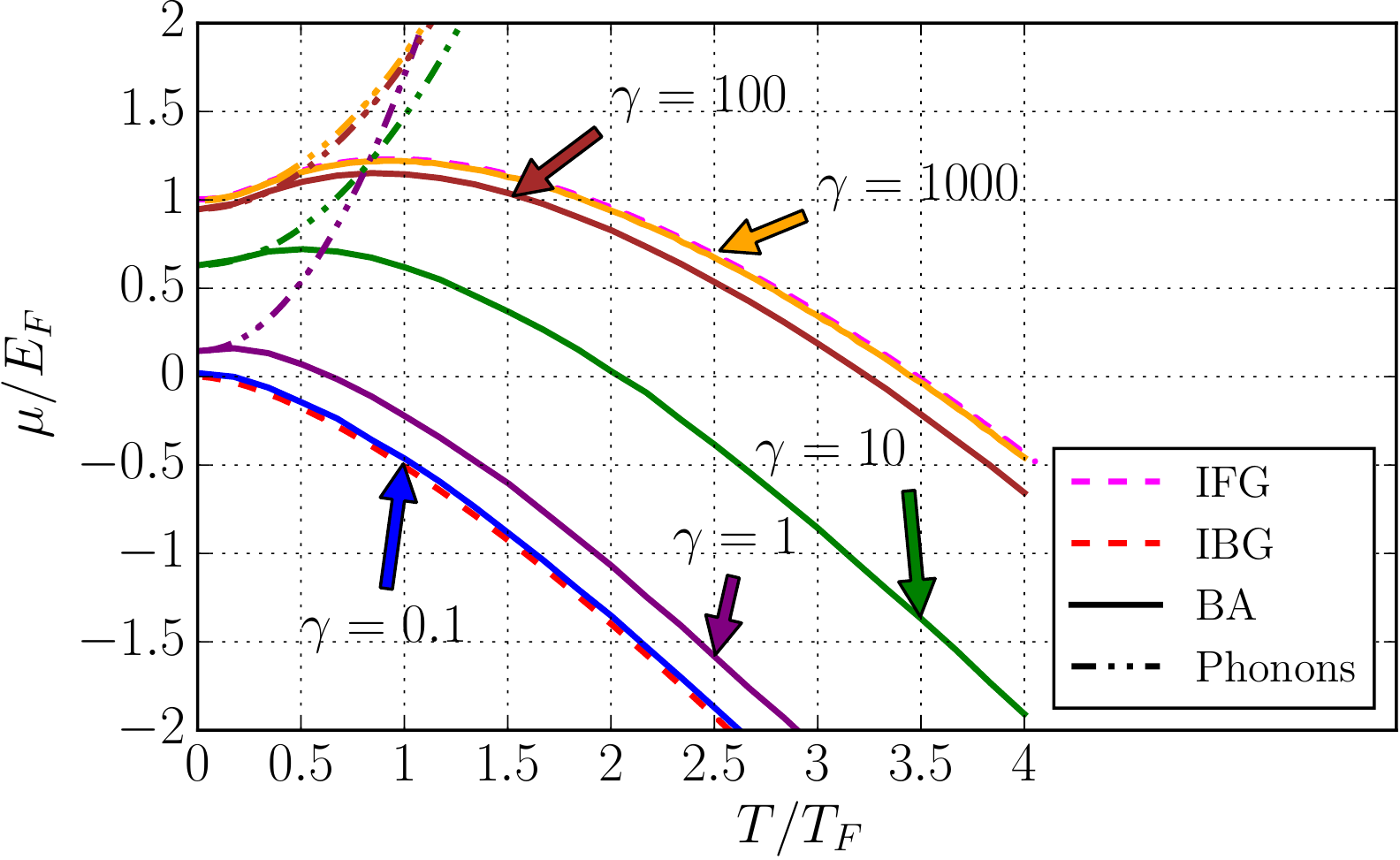}
\caption{(Color online) Chemical potential as a function of temperature $T$ in Fermi units for several values of $\gamma$ and at a fixed density $n|a_{\rm 1D}| = 2/\gamma$. The solid lines represent the Bethe--{\it ansatz} (BA) solutions for different values of $\gamma$. The dot-dashed lines are the low--temperature expansions of the chemical potential taking into account only the phononic contribution, Eq.~\eqref{Eq:mu general}. The phononic expansions for $\gamma \geq 1000$ are equal to the analytical Sommerfeld expansion of Eq.~\eqref{Eq:Sommerfeld}. Both the chemical potentials as a function of $T$ for the ideal Fermi (upper dashed line) and ideal Bose (lower dashed line) gas are also reported, Eq.~\eqref{Eq:FD} and~\eqref{Eq:IBG}, respectively. \label{fig:comparison}}
\end{figure}

\begin{figure}[ht]
\centering
\includegraphics[scale=0.5]{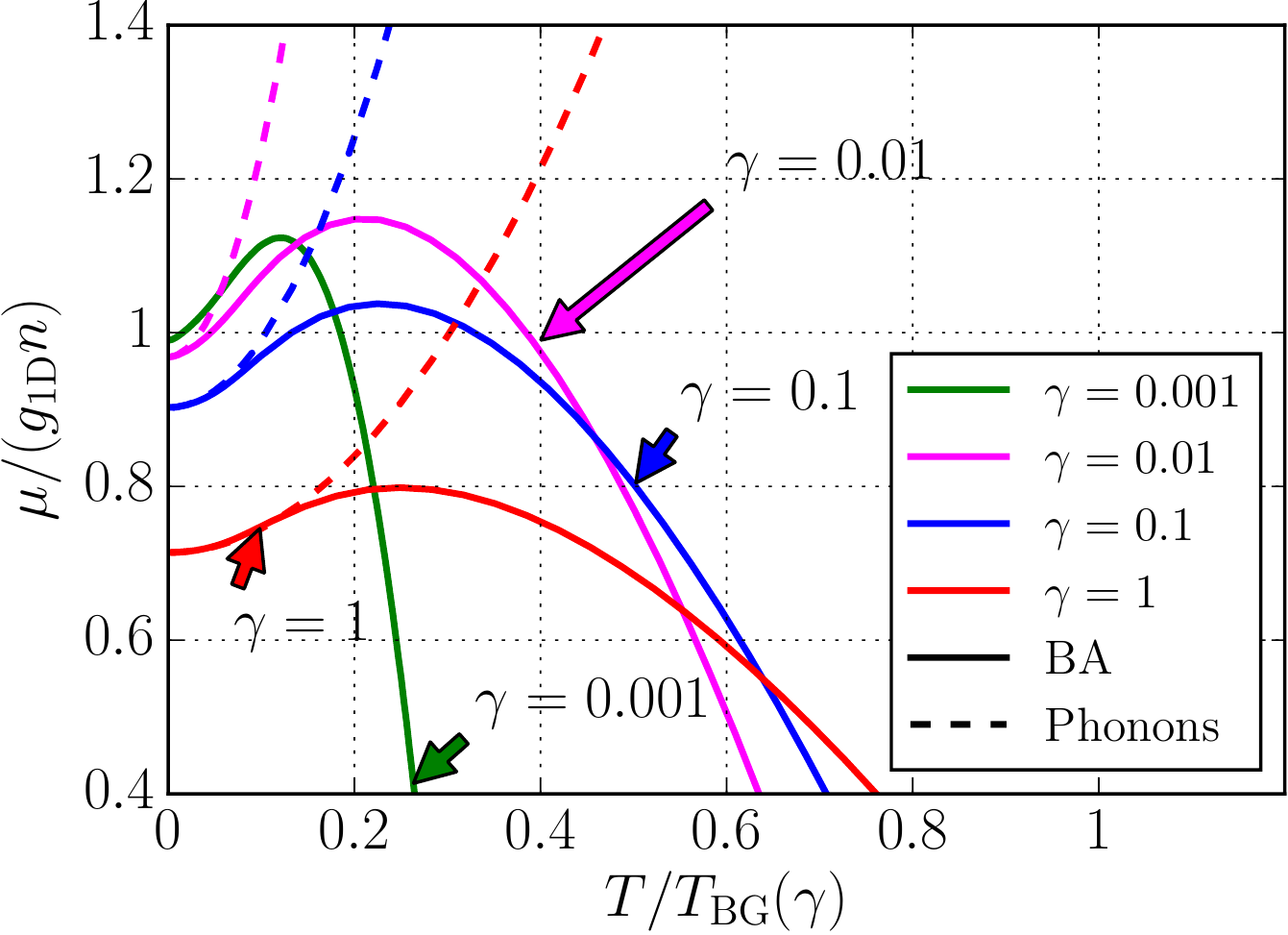}
\caption{(Color online) Chemical potential as a function of temperature in BG units for several values of $\gamma$. The solid lines represent the Bethe--{\it ansatz} (BA) solutions for different values of $\gamma$. The dashed lines are the low--temperature expansions of the chemical potential taking into account only the phononic contribution, Eq.~\eqref{Eq:mu_general_gn}. We notice also that the phononic expansion~\eqref{Eq:mu_general_gn} does not hold for very small $\gamma$, like $\gamma = 0.001$, for which value the low--temperature expansion is not reported. \label{fig:mu_gn}}
\end{figure}

In Figs.~\ref{fig:comparison}-\ref{fig:mu_gn} we report the temperature dependence of the chemical potential of the system described by Hamiltonian~\eqref{Eq:Hamilt} as obtained numerically from the Bethe-{\it ansatz} (BA) approach first developed by Yang-Yang~\cite{Yang1969, Yang1970, Kheruntsyan2005, Lang2015} for several characteristic values of $\gamma$. The Yang-Yang description has been probed experimentally \cite{vanAmerongen2008, Vogler2013} and allows not only to investigate the thermodynamics, but also the Luttinger liquid physics and the quantum criticality of the system \cite{Guan2011, Wang2013, Guan2014}.
The numerical results for the thermodynamics have been
derived recently in an analytic fashion by using the polylog
functions at finite temperature by Guan and Batchelor \cite{Guan2011, Guan2014}.

The crossover from mean-field to Tonks-Girardeau regimes (see Fig.~\ref{fig:v}) introduces two distinct energy scales.
Correspondingly, we rescale the chemical potential in units of the Fermi energy $E_F$ in Fig.~\ref{fig:comparison} and in units of the mean-field zero-temperature chemical potential $\mu_{\rm BG}(T = 0) = g_{\rm 1D}n$ in Fig.~\ref{fig:mu_gn}.
The first choice provides natural units in the TG regime in which strongly repulsive bosons behave similarly to an ideal Fermi gas (IFG) in the limit of $\gamma\to\infty$. In this regime, the chemical potential as a function of $T$ is calculated by inverting the Fermi--Dirac distribution (upper dashed line in Fig.~\ref{fig:comparison}):
\begin{equation}
\label{Eq:FD}
n_{\rm IFG}(p) = \frac{1}{e^{\frac{1}{k_BT}\left(\frac{p^2}{2m} - \mu\right)} + 1}\ ;
\end{equation}
and, despite the absence of superfluidity, it still exhibits the quadratic low-temperature dependence $\mu \propto T^2$, which follows from the low-temperature Sommerfeld expansion, Eq.~\eqref{Eq:Sommerfeld}.

By reducing the interaction parameter $\gamma$, the system becomes softer and the limit of vanishing interactions, $\gamma\to 0 $, corresponds to an ideal Bose gas (IBG) with the chemical potential $\mu(T)$ fixed by the relationship (lower dashed line in Fig.~\ref{fig:comparison}):
\begin{equation}
\label{Eq:IBG}
n_{\rm IBG}(p) = \frac{1}{e^{\frac{1}{k_BT}\left(\frac{p^2}{2m} - \mu\right)} - 1}\;.
\end{equation}
Notice that, because of the absence of Bose-Einstein condensation~\cite{Mermin1966,Hohenberg1967}, the chemical potential of the 1D ideal Bose gas is always negative and approaches the value $\mu=0$ as $T \to 0$.
Remarkably, for all finite interaction strengths the temperature dependence is not monotonic.
Moreover, the initial increase is perfectly described by the quadratic low-temperature expansion~(\ref{Eq:mu general}), thereby proving that the model based on a gas of independent phonons well accounts for the thermodynamic behaviour of the 1D interacting Bose gas. This is a non trivial result due to the complex structure of the elementary excitations at larger wave vectors exhibiting a double branch converging into the phonon law~(\ref{Eq:vsp}) only at small momenta.
We notice also that the chemical potential for high temperatures, which is a decreasing function of $T$, can be considered as a shift of the ideal Bose chemical potential, Eq.~\eqref{Eq:IBG}, for every value of $\gamma$.

The behavior of the chemical potential in the weakly-interacting regime ($\gamma \ll 1$) is best seen in Fig.~\ref{fig:mu_gn}.
For low temperatures $T \ll \mu$  the gas behaves like a quasicondensate, exhibiting typical features of superfluids. For $\mu \ll T \ll T_F$, the gas is a thermal degenerate gas, while for $T \gg T_F$ the gas behaves classically with $\mu < 0$.
A similar classification of the quantum degeneracy states in 1D trapped configurations was first proposed in~\cite{Petrov2000}.

Although there is no phase transition in 1D systems at finite $T$, in the canonical ensemble, there exists a
critical point, corresponding to the value $\mu = 0$ of  the chemical potential, which separates
the vacuum from the filled ``Fermi sea''  of repulsive bosons
at $T = 0$. In particular, a universality class is present in the temperature regime $T \gg |\mu|$ and near the critical point $\mu = 0$ \cite{Guan2011, Wang2013, Guan2014}.

Figure~\ref{fig:mu_gn} is similar to Fig.~\ref{fig:comparison}, but with the chemical potential expressed in units of the BG chemical potential at zero temperature: $\mu_{\rm BG}(T = 0) = g_{1D}n$ and the temperature in units of:
\begin{equation}
\label{Eq:TBG}
T_{\rm BG}(\gamma) =  \frac{mv_F^2 \sqrt{\gamma}}{\pi k_B}
\end{equation}
which has been introduced as an appropriate temperature scale for visualizing the behavior of the chemical potential at low temperature.
With the new units, the phononic expansion~\eqref{Eq:mu general} takes the form:
\begin{equation}
\label{Eq:mu_general_gn}
\mu(T, \gamma) = g_{\rm 1D}n \left[ \alpha(\gamma) \frac{\pi^2}{2\gamma} + 2\beta(\gamma) \left(\frac{T}{T_{\rm BG}} \right)^2 \right]\ .
\end{equation}

Figures~\ref{fig:comparison} and~\ref{fig:mu_gn} point out in a clear way the non-monotonic behavior of the chemical potential $\mu$ as a function of $T$ for a fixed value of the density. 
This is a general feature exhibited by superfluids \cite{Papoular2012} and it is shown here that it characterizes also interacting 1D Bose gas for all finite values of the interaction parameter $\gamma$.

Both figures show also that the phononic expansion describes very well
the low-T thermodynamics for all values of $\gamma$, although the
region of the applicability of the phononic description depends on
$\gamma$. As pointed out in Ref.~\cite{Lang2017}, for small values of
the interaction parameter $\gamma$, higher-order corrections beyond
the linear phononic contribution in the excitation
spectrum~\eqref{Eq:vsp} might be important.

\section{Bogoliubov regime $\gamma \to 0$}
\label{Sec:BOG}


In the mean-field theory, the chemical potential is linear in density, $\mu_{\rm BG}(T = 0) = g_{\rm 1D}n$ and the velocity of sound takes the value $v_s^{\rm BG}(\gamma) = \hbar n \sqrt{\gamma}/m = v_F \sqrt{\gamma}/\pi$, see Fig.~\ref{fig:v}.

The first correction to the mean-field expression for the equation of state comes from the quantum fluctuations~\cite{Lee1957, Lee19572, Pitaevskii2016}. With respect to the 3D case, in 1D this calculation is simpler because it does not require the renormalization of the scattering length due to the absence of ultraviolet divergencies in the calculation of the ground-state energy. Therefore in 1D one can consider all ranges of momenta and one finds~\cite{Lieb1963}:
\begin{equation}
\frac{E_0}{N} = \frac{1}{2}g_{\rm 1D}n + \frac{2}{2N}\sum_{p > 0}^{+\infty} \left[\epsilon(p) - g_{\rm 1D}n - \frac{p^2}{2m} \right]
\label{Eq:beyondBG}
\end{equation}
where
\begin{equation}
\label{Eq:BG spectrum}
\epsilon(p) = \sqrt{\frac{g_{\rm 1D}n}{m}p^2 + \left(\frac{p^2}{2m} \right)^2}
\end{equation}
is the Bogoliubov excitation spectrum.
By considering the thermodynamic limit of Eq.~\eqref{Eq:beyondBG} and by solving the integral in momentum space, one finally finds the first-order correction in the interaction parameter for the ground state energy~\cite{Lieb1963}
\begin{equation}
\label{Eq:E BG corr}
\frac{E_0}{N}(\gamma \ll 1) = \frac{\hbar^2 n^2}{2m}\gamma \left(1 - \frac{4}{3\pi} \sqrt{\gamma} \right) \ .
\end{equation}
The same result can be also found by performing a power series expansion of the Lieb-Liniger equations~\cite{Kaminaka2011, Gudyma20152}.
The correction is negative as it comes from second order perturbation theory and, contrary to the higher dimensions, in 1D there is no renormalization of the coupling constant thus no additional terms have to be added.

Equation~\eqref{Eq:E BG corr} allows one to calculate the higher-order corrections for the other thermodynamic quantities at $T = 0$. For the chemical potential, one finds the result
\begin{equation}
\label{Eq:mu BG corr}
\mu(\gamma \ll 1) \approx \frac{\hbar^2 n^2 \gamma}{m} \left(1 - \frac{\sqrt{\gamma}}{\pi} \right)
\end{equation}
which implies the result
\begin{equation}
\label{Eq:alpha BG corr}
\alpha(\gamma \ll 1) \approx \alpha_{\rm BG}(\gamma)\left[1 - \frac{\sqrt{\gamma}}{\pi}\right]
\end{equation}
for the expansion of the coefficient $\alpha(\gamma)$, where $\alpha_{\rm BG}(\gamma) = 2\gamma/\pi^2$ is the mean-field value. The corresponding result has been plotted in Fig.~\ref{fig:alphacorr} and well reproduces the exact value of $\alpha(\gamma)$ up to values $\gamma \sim 1$.

From Eq.~\eqref{Eq:vs} and Eq.~\eqref{Eq:mu BG corr}, one can calculate also the correction to the sound velocity~\cite{Lieb1963}
\begin{equation}
\label{Eq:vs BG corr}
v_s(\gamma \ll 1) \approx v_s^{\rm BG}(\gamma) \sqrt{1 - \frac{\sqrt{\gamma}}{2\pi}}
\end{equation}
which is also reported in Fig.~\ref{fig:v}, yielding the expression
\begin{equation}
\label{Eq:beta BG corr}
\beta(\gamma \ll 1) \approx \beta_{\rm BG}(\gamma)\ ,
\end{equation}
for the coefficient $\beta(\gamma)$, Eq.~\eqref{Eq:beta}
with $\beta_{\rm BG}(\gamma) = \pi^3/(24\sqrt{\gamma} )$ the Bogoliubov value. Notice that, differently from the case of $\alpha(\gamma)$ [see Eq.~\eqref{Eq:alpha BG corr}], the first correction $\beta_{\rm BG}(\gamma)$ vanishes because of an exact cancellation between the corrections provided by the terms $\partial v_s/\partial n$ and $v_s^2$ of Eq.~\eqref{Eq:beta}. This explains why the Bogoliubov approximation describes correctly the value of $\beta(\gamma)$ for a large interval of values of $\gamma$, up to $\gamma \sim 1$ (see Fig.~\ref{fig:beta}).


\section{Tonks-Girardeau regime $\gamma \rightarrow \infty$}
\label{Sec:TG}

In the TG limit of strong repulsion, $\gamma \rightarrow \infty$, the energetic properties are the same as in an ideal Fermi gas. 
The thermodynamic quantities do not depend on the coupling constant $g_{\rm 1D}$, but only on the density $n$, encoded in the Fermi energy $E_F$. 
This regime can be interpreted as that of a unitary Bose gas with the Bertsch parameter equal to 1 as the chemical potential is equal to the Fermi energy [$\mu_{\rm TG}(T = 0) = E_F$].  
Similarly, the sound velocity is equal to the Fermi velocity $v_s^{\rm TG} = v_F = \sqrt{2E_F/m}$, see Fig.~\ref{fig:v}.
The low-temperature expansion of the chemical potential in this limit is equal to the first terms of the Sommerfeld expansion (dot-dashed line for $\gamma = 1000$ in Fig.~\ref{fig:comparison}) of the 1D ideal Fermi gas, as already pointed out in~\cite{Lang2015}:
\begin{equation}
\label{Eq:Sommerfeld}
\mu_{Somm}(T) = E_F\left[1 + \frac{\pi^2}{12}\left(\frac{T}{T_F}\right)^2 \right]
\end{equation}
which contains the TG limits of $\alpha(\gamma)$ and $\beta(\gamma)$ parameters, Figs.~\ref{fig:alphacorr} and \ref{fig:beta}.

Leading corrections to the ground-state energy in the TG regime arise from the ``excluded volume'' and can be obtained from the equation of state of hard spheres (i. e. impenetrable) bosons with diameter $a_{\rm 1D} > 0$~\cite{Girardeau1960}:
\begin{equation}
\label{Eq:E hard sphere}
\frac{E_0}{N} = \frac{\pi^2\hbar^2}{6m}\frac{n^2}{(1 - na_{\rm 1D})^2} \ .
\end{equation}
In the limit of point-like bosons $a_{\rm 1D} = 0$, Eq.~\eqref{Eq:E hard sphere} reproduces the ground state energy of the ideal Fermi gas, $E_{\rm TG} = \pi^2 \hbar^2 n^2/(6m)$.
Expanding the denominator in Eq.~(\ref{Eq:E hard sphere}) generates a power series with integer coefficients,
$E / E_{\rm TG} = 1 + 2na_{\rm 1D} + 3(na_{\rm 1D})^2 + 4(na_{\rm 1D})^3 + \cdots$.
It is interesting to notice that for a $\delta$-interacting potential the momentum-dependent $s$-wave scattering length, $a_{\rm 1D}(k)  = \arctan(ka_{\rm 1D})/k = a_{\rm 1D} - (1/3) k^2a_{\rm 1D}^2$, does not affect first and second corrections in $na_{\rm 1D}$ but induces a negative correction in front of the third correction.
Indeed, the universality of the first and the second corrections becomes evident by comparing low-density expansion of the equation of state for hard spheres, Eq.~(\ref{Eq:E hard sphere}), and contact $\delta$-potential obtained by solving Bethe equations recursively~\cite{Astrakharchik2010},
\begin{equation}
\label{Eq:E corr series}
\frac{E_0}{N}\!=\!\frac{\pi^2\hbar^2n^2}{6m}\left[1\!+\!2na_{\rm 1D}\!+\!3(na_{\rm 1D})^2\!+\!\left(4-\frac{4\pi^2}{15}\right)(na_{\rm 1D})^3
\right] \ .
\end{equation}
The non-universal correction depends on the shape of the potential and for the LL model it has a non-integer coefficient, which qualitatively can be understood by noting that the typical value of the scattering momentum in TG regime is proportional to $k_F = \pi \hbar n / m$, which is consistent with $\pi^2$ terms appearing in expansion~(\ref{Eq:E corr series}).
The universal terms are the same both in the super Tonks-Girardeau~\cite{Astrakharchik2005} 
($a_{\rm 1D} > 0$) and the strongly repulsive~\cite{Gudyma20152} ($a_{\rm 1D} < 0$) regimes.
From Eq.~\eqref{Eq:E corr series}, by introducing the parameter~\eqref{Eq:gamma} and by considering only the leading term, one finds
\begin{equation}
\label{Eq:E corr big int}
\frac{E_0}{N}(|\gamma| \gg 1) \approx \frac{\pi^2 \hbar^2 n^2}{6m}\left(1 - \frac{4}{\gamma} \right) \ .
\end{equation}

From Eq.~\eqref{Eq:E corr big int}, one easily calculates the correction of the chemical potential at $T = 0$, Eq.~\eqref{Eq:alpha}:
\begin{equation}
\label{Eq:mu corr big int}
\mu(|\gamma| \gg 1) \approx E_F \left(1 - \frac{16}{3\gamma} \right)
\end{equation}
which implies the result~\cite{Gudyma20152}
\begin{equation}
\label{Eq:alpha corr big int}
\alpha(|\gamma| \gg 1) \approx 1 - \frac{16}{3\gamma}
\end{equation}
for $\alpha(\gamma)$, including the first correction to the TG result $\alpha_{\rm TG} = 1$.
Prediction~(\ref{Eq:alpha corr big int}) is reported in Fig.~\ref{fig:alphacorr} for positive values of $\gamma $, its accuracy being good for values of $\gamma$ larger than $\sim 10$.

From Eq.~\eqref{Eq:vs} and Eq.~\eqref{Eq:mu corr big int}, one can calculate also the first correction, at large $\gamma$, to the sound velocity~\cite{Gudyma20152, Valiente2016}:
\begin{equation}
\label{Eq:vs corr big int}
v_s(|\gamma| \gg 1) \approx v_F \sqrt{1 - \frac{8}{\gamma}} \;
\end{equation}
which is reported in Fig.~\ref{fig:v}. For the coefficient $\beta(\gamma)$, which provides the $T^2$-correction in the expansion of the chemical potential, we find again an exact cancellation between the $1/\gamma$ correction (provided by the term $\partial v_s/\partial n)$ and $v_s^2$ entering the expression~\eqref{Eq:beta} for $\beta(\gamma)$, similarly to what happens in the small $\gamma$ expansion discussed in the previous Section \ref{Sec:BOG} in the case of the Bogoliubov gas. We then find that the Tonks-Girardeau expression $\beta_{\rm TG} = \pi^2/12$ provides an accurate estimate of $\beta(\gamma)$ for values of $\gamma$ larger than $\sim 10$ (see Fig.~\ref{fig:beta}).

More accurate analytical expressions for the above thermodynamical quantities, which allow to probe the whole range of interaction strength with excellent accuracy, are reported in \cite{Ristivojevic2014, Jiang2015, Prolhac2017, Lang2017}.

\section{Low-temperature expansion of the inverse compressibility}
\label{Sec:compress}

Here we derive the dependence of the adiabatic and isothermal inverse compressibilities on the interaction parameter $\gamma$ in the limit of low temperature.

\subsection{Adiabatic inverse compressibility and sound velocity}
\label{SubSec:adiabatic compressibility}
From the Gibbs--Duhem relation $dP = nd\mu + sdT$, one finds
\begin{equation}
\label{Eq:adiabatic compress}
\left(\frac{\partial P}{\partial n} \right)_{\bar{s}} = n \left( \frac{\partial \mu}{\partial n}\right)_{\bar{ s}} + n\bar{s}\left(\frac{\partial T}{\partial n}\right)_{\bar{s}}
\end{equation}
where $s$ is  the entropy density and $\bar{s}=s/n$ is the entropy per particle.

\begin{figure}[h!]
\centering
\includegraphics[scale=0.55]{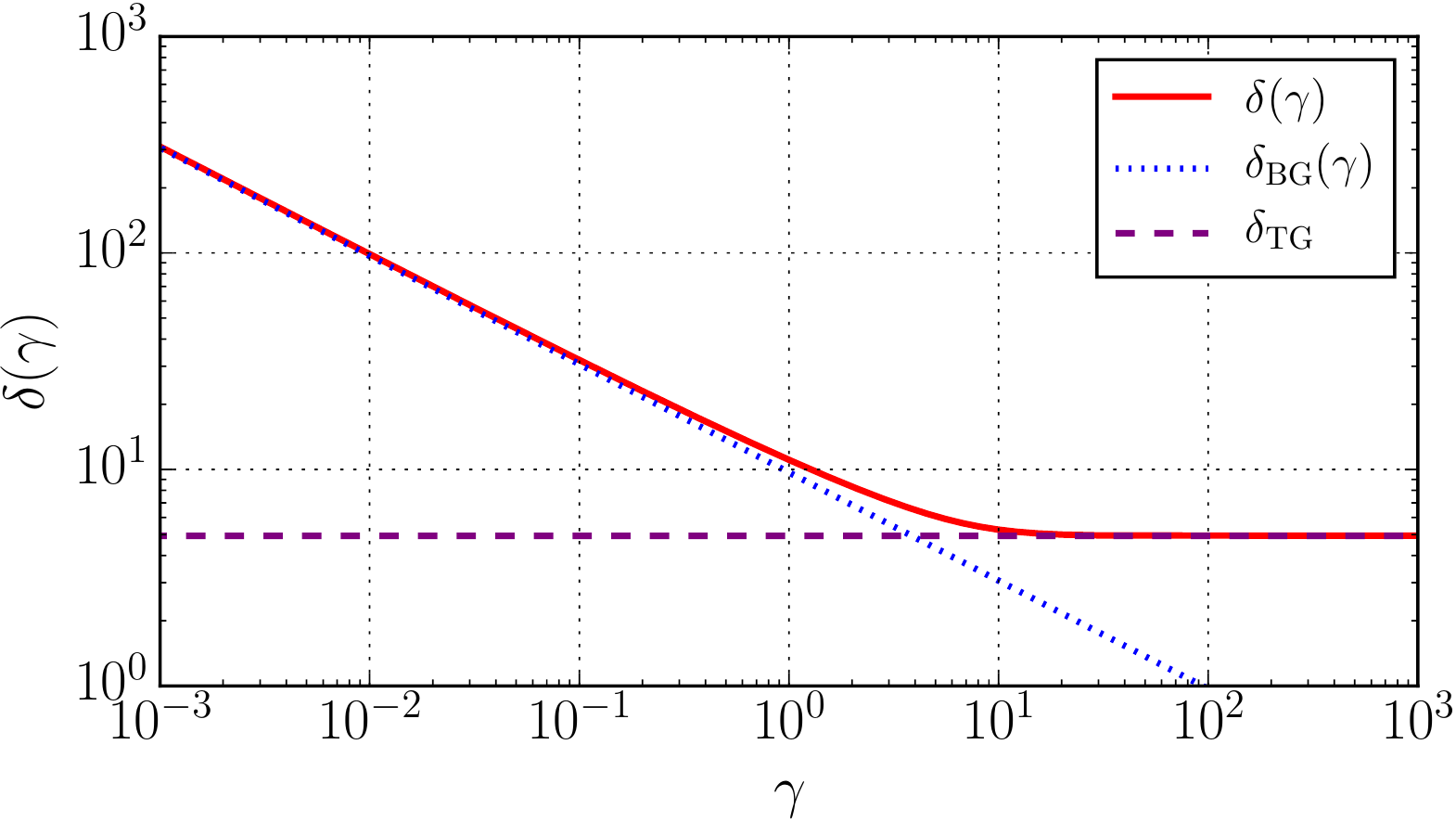}
\caption{(Color online) Value of the dimensionless coefficient $\delta(\gamma)$ of the low--temperature expansion of the adiabatic inverse compressibility (solid line). The BG and TG analytical limits are also shown: $\delta_{\rm BG}(\gamma) =  5 \pi^3/(16 \sqrt{\gamma})$ (dotted line) and $\delta_{\rm TG} = \pi^2/2$ (dashed line). \label{fig:delta}}
\end{figure}

At low temperature the entropy per particle of a non--interacting gas of phonons takes the form~\cite{Pitaevskii2016}
\begin{equation}
\label{Eq:sbar}
\bar{s}(T)=  \frac{\pi k_B^2 T}{3\hbar v_s n} \ ,
\end{equation}
which depends on the $T=0$ value~\eqref{Eq:vs} of the sound velocity. 
Use of relation~\eqref{Eq:sbar} permits to express the dependence of the second contribution to the adiabatic inverse compressibility on the r.h.s. of Eq.~(\ref{Eq:adiabatic compress}) on the interaction parameter $\gamma$
\begin{equation}
\label{Eq:T n}
\left(\frac{\partial T}{\partial n}\right)_{\bar{s}} = \frac{3\hbar v_s \bar{s}}{\pi k_B^2}\left(1 + \frac{6\hbar n v_s \beta(\gamma)}{\pi E_F}  \right) \ ,
\end{equation}
in terms of the coefficient $\beta(\gamma)$, Eq.~\eqref{Eq:beta} related to the density derivative of the sound velocity at constant entropy. 
The first contribution on the r.h.s. of Eq.~(\ref{Eq:adiabatic compress}) can be obtained by using Eqs.~\eqref{Eq:mu general} and~\eqref{Eq:sbar},
\begin{equation}
\label{Eq:mu n}
\left(\frac{\partial \mu}{\partial n}\right)_{\bar{s}} = \frac{m}{n}v_s^2  + \frac{ (k_BT)^2}{nE_F}\left(\frac{12n \hbar v_s}{\pi E_F}\beta^2(\gamma) - \gamma \frac{\partial \beta(\gamma)}{\partial \gamma}  \right) \, .
\end{equation}
From the above equations one finally finds the low temperature expansion
\begin{equation}
\label{Eq:P n sbar}
\left(\frac{\partial P(T, \gamma)}{\partial n}\right)_{\bar{s}} = \left(\frac{\partial P(\gamma)}{\partial n}\right)_{T=0}  + E_F \delta(\gamma)\left( \frac{T}{T_F}\right)^2
\end{equation}
of the adiabatic inverse compressibility, where 
\begin{equation}
\left(\frac{\partial P(\gamma)}{\partial n}\right)_{T=0} = m v_s^2(\gamma)
\end{equation}
is its $T=0$ value and we have defined the positive quantity
\begin{equation}
\label{Eq:delta}
\delta(\gamma) = \frac{24}{\pi^2}\beta^2(\gamma) \frac{v_s(\gamma)}{v_F} - \gamma \frac{\partial \beta(\gamma)}{\partial\gamma} + \frac{\pi^2}{6}\frac{v_F}{v_s(\gamma)} + 2\beta(\gamma) \ ,
\end{equation}
which is reported in Fig.~\ref{fig:delta} together with its asymptotic limits in the Bogoliubov and Tonks--Girardeau regimes.

\subsection{Isothermal inverse compressibility}
\label{SubSec:isothermal compress}

By fixing the temperature $T$ in Eq.~\eqref{Eq:adiabatic compress} and by considering the low-temperature expansion of the chemical potential~\eqref{Eq:mu general}, one can also calculate the low-temperature expression for the isothermal inverse compressibility
\begin{equation}
\label{Eq:explicit isothermal compress}
\left(\frac{\partial P(T, \gamma)}{\partial n}\right)_T = \left(\frac{\partial P(\gamma)}{\partial n}\right)_{T=0}  +  E_F \eta(\gamma) \left(\frac{T}{T_F} \right)^2
\end{equation}
where we have defined the negative dimensionless coefficient
\begin{equation}
\label{Eq:eta}
\eta(\gamma) = - 2\beta(\gamma) - \gamma \frac{\partial \beta(\gamma)}{\partial \gamma} \ .
\end{equation}

\begin{figure}[htbp]
\centering
\includegraphics[scale=0.55]{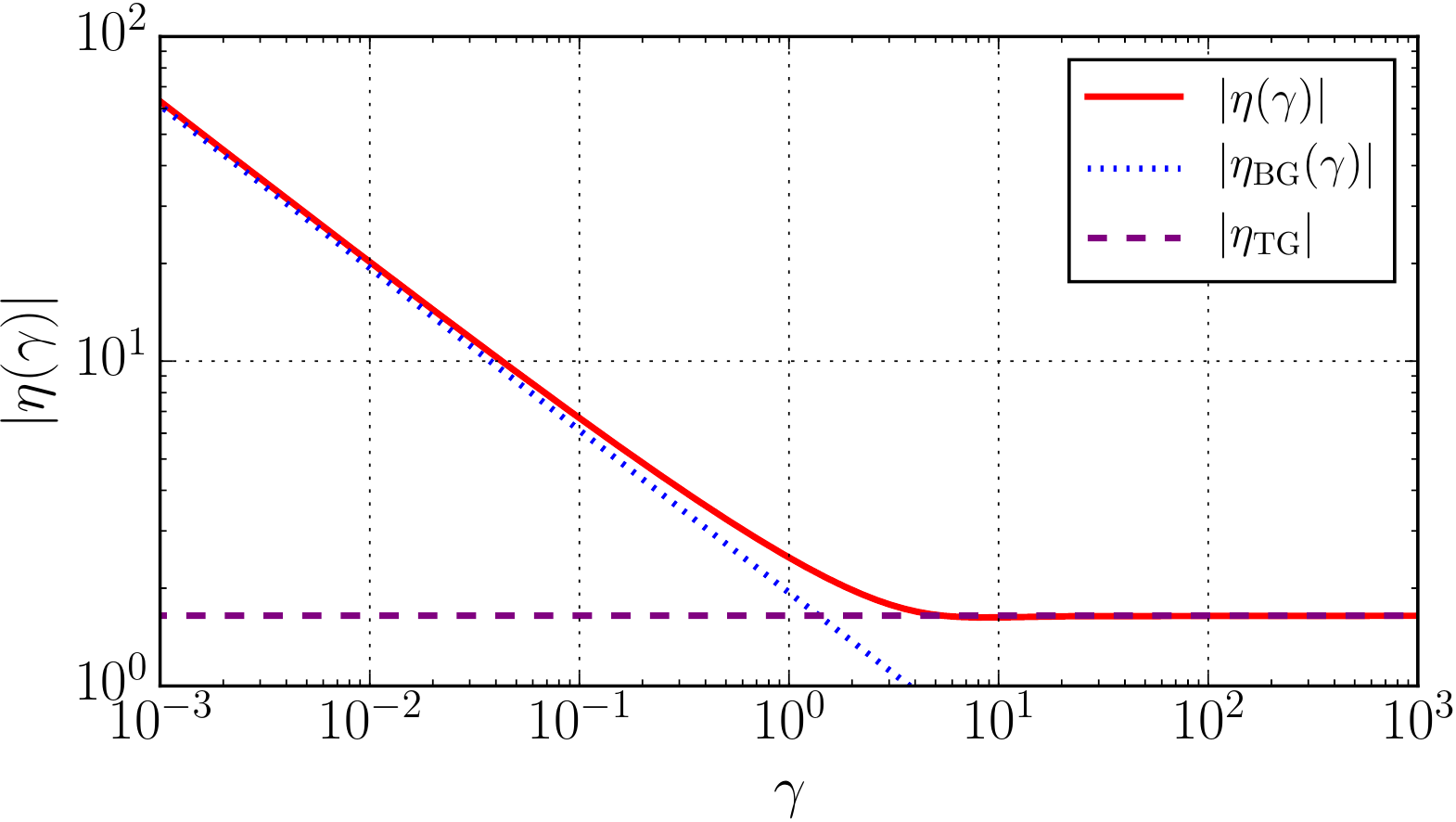}
\caption{(Color online) Absolute value of the dimensionless coefficient $\eta(\gamma)$ of the low--temperature expansion of the isothermal inverse compressibility (solid line). The BG and TG analytical limits are also shown: $\eta_{\rm BG}(\gamma) = - \pi^3/(16 \sqrt{\gamma})$ (dotted line) and $\eta_{\rm TG} =  - \pi^2/6$ (dashed line). \label{fig:eta}}
\end{figure}

Notice that the thermal corrections to the isothermal and adiabatic inverse compressibilities have opposite sign, being the coefficient $\eta(\gamma)$ always negative. The absolute value of $\eta(\gamma)$ is reported in Fig.~\ref{fig:eta} together with the asymptotic limits in the Bogoliubov and Tonks--Girardeau regimes. The negative value of $\eta(\gamma)$ is the consequence of the peculiar temperature dependence of the free energy \eqref{Eq:A_lowT}.

\section{Gas on a ring}
\label{Sec:pbc}

The physics in one dimension is unusual in many aspects.
The mean--field regime is reached at {\em large} densities contrarily to what happens in three dimensions where the weakly--interacting limit corresponds to small densities, according to the limit $na^3\to 0$.
For a fixed number of particles $N$ the mean--field limit in one dimension, $n|a_{1D}|\to\infty$, can be obtained either increasing the linear density $n = N/L$, by decreasing the system size $L$, or by increasing the $s$--wave scattering length $a_{1D}$, i.e. decreasing the coupling constant $g_{1D} =
-2\hbar^2/(ma_{1D})$.
Asymptotically, at a certain point, the size of the system $L$ will become comparable to the healing length
\begin{equation}
\label{Eq:healing}
\xi = \sqrt{\frac{\hbar^2}{2mg_{\rm 1D}n}}
\end{equation}
and finite--size effects will become important.
This should be contrasted to the three-dimensional case where the mean--field regime is instead  achieved by increasing the system size $L$ which consequently becomes larger than the healing length.

Finite--size effects depend on the system geometry.
Interestingly, periodic boundary conditions, commonly used as a mathematical tool in the three--dimensional world, in one dimension can be explicitly realized in a ring and have consequently a direct
physical interest.
This is another peculiarity of the one--dimensional world.
In the following we calculate the finite--size dependence of thermodynamic quantities for a gas confined in a ring whose properties are then equivalent to the ones of a linear 1D system satisfying
periodic boundary conditions (PBC). 
If one considers a plane wave $\propto e^{ikz}$ and one imposes PBC, one finds that the momentum is quantized according to
\begin{equation}
\label{Eq:quantized p}
p_i = \hbar k_i = \frac{2\pi \hbar n_i}{L}
\end{equation}
where $n_i = 0, \pm$ are integers.
Moreover, in 1D, all the integrals in momentum space, defined in the thermodynamic limit ($N, L \rightarrow + \infty$, $n =$ finite), are replaced by a sum over the discretized momenta~\eqref{Eq:quantized p} as:
\begin{equation}
\label{Eq:sum integral p}
\int_{-\infty}^{+\infty}dp \to  \frac{2\pi \hbar}{L}\sum_{p = -\infty}^{+\infty} \ .
\end{equation}

In the following, we calculate the finite-size corrections in both BG and TG regimes at zero temperature, as well as the static inelastic structure factor for a finite number of particles. 

\subsection{Bogoliubov regime at $T = 0$}

Let us consider the $T = 0$ ground-state energy per particle given by
\begin{equation}
\label{Eq:E no self}
\frac{E_0}{N} = \frac{1}{2}g_{\rm 1D}n + \frac{1}{2N} \sum_{p = -\infty}^{+ \infty} \left[ \epsilon(p) - g_{\rm 1D}n - \frac{p^2}{2m} \right]
\end{equation}
corresponding to the Bogoliubov regime of small $\gamma$, where $\epsilon(p)$ is provided by the Bogoliubov spectrum~\eqref{Eq:BG spectrum}.
Equation~\eqref{Eq:E no self} differs from Eq.~\eqref{Eq:beyondBG} because it contains the $p = 0$ term in the sum. This term has been included in order to avoid self-interaction effects in the leading mean-field term of Eq.~\eqref{Eq:beyondBG} which should be replaced by $g_{\rm 1D}(N - 1)/(2L)$.

By introducing the discretized values of $p$~\eqref{Eq:quantized p}, the energy can be rewritten in the form
\begin{equation}
\label{Eq:E G}
\frac{E_0}{N} = \frac{1}{2}g_{\rm 1D}n \left[1 + \sqrt{\gamma} G(y) \right]
\end{equation}
where we have introduced the dimensionless variable
\begin{equation}
\label{Eq:y}
y = \gamma N^2,
\end{equation}
depending on the interaction parameter $\gamma$ and the function
\begin{equation}
\label{Eq:G}
G(y) = \frac{2}{y\sqrt{y}}\sum_{n_i = 0}^{+\infty} \left[  2 \pi n_i \sqrt{y + (\pi n_i)^2} - 2(\pi n_i)^2 - y\right] + \frac{1}{\sqrt{y}} \ ,
\end{equation}
where the adding of the quantity $1/\sqrt{y}$ ensures that the term $n_i = 0$ in the sum is counted just once.

\begin{figure}[htbp]
\centering
\includegraphics[scale=0.5]{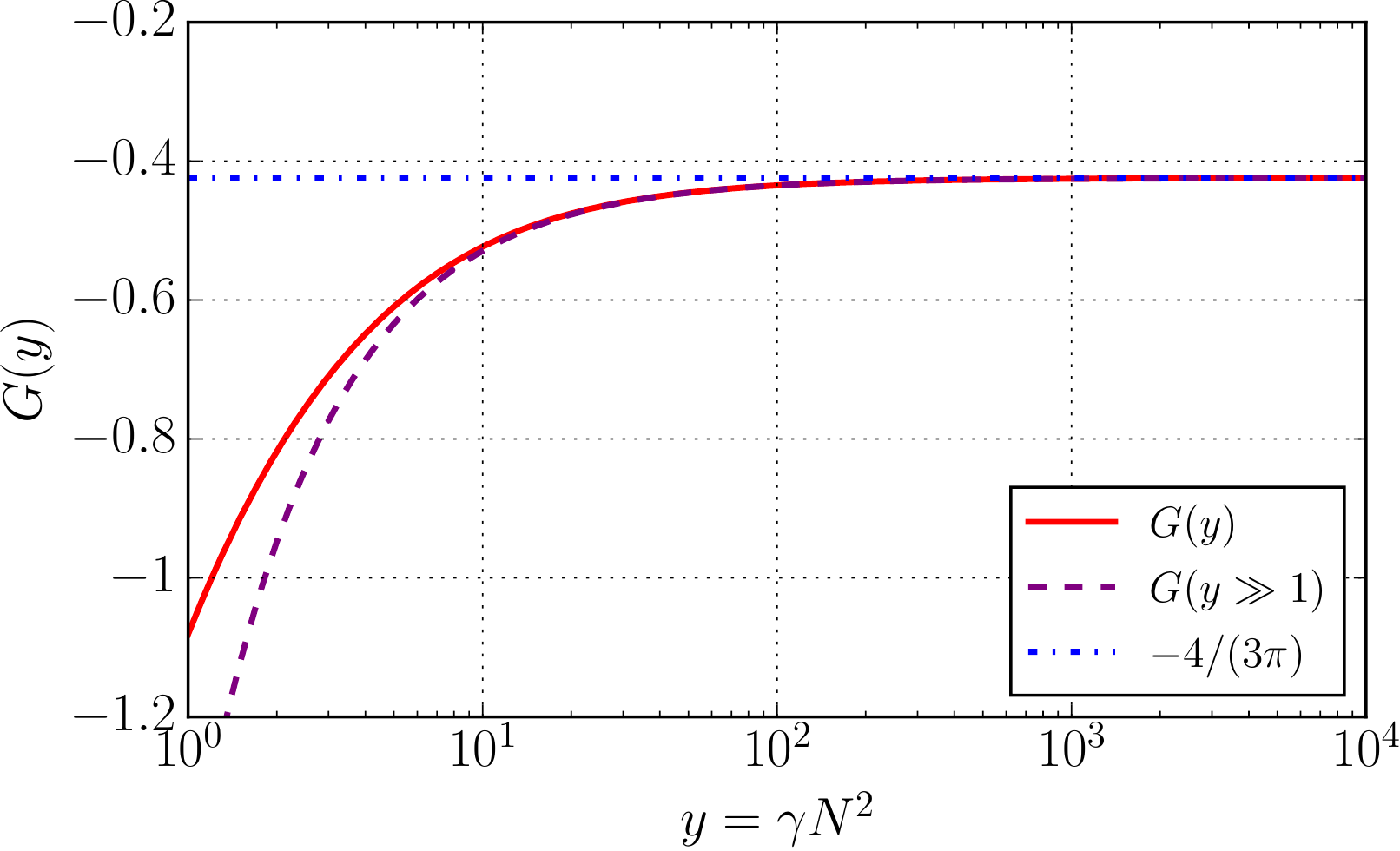}
\caption{(Color online) Comparison of the numerical series $G(y)$~\eqref{Eq:G} (solid line) and its analytical expansion~\eqref{Eq:GL} (dashed line) holding for $y \gg 1$. The dot-dashed line represents the thermodynamic value. \label{fig:G}}
\end{figure}

By using the Euler-Maclaurin expansion (see Appendix~\ref{Sec:G}), one can calculate the expression for the series~\eqref{Eq:G} for large values of $y$:
\begin{equation}
\label{Eq:GL}
G(y \gg 1) \approx - \frac{4}{3 \pi} - \frac{\pi}{3y} \ .
\end{equation}
In Fig.~\ref{fig:G} we report the comparison of the series~\eqref{Eq:G} with its expansion~\eqref{Eq:GL}. We notice that the two curves agree in an excellent way for $y > 10$. The thermodynamic limit $-4/(3\pi)$ is also reported.

For large number of particles, the ground-state energy per particle~\eqref{Eq:E G} then takes the form:
\begin{equation}
\label{Eq:E BG ring limit}
\frac{E_0}{N}(\gamma N^2 \gg 1, \gamma \ll 1) \approx \frac{1}{2}g_{\rm 1D}n \left[1 - \frac{4}{3 \pi} \sqrt{\gamma} - \frac{\pi}{3N^2 \sqrt{\gamma}} \right]
\end{equation}
and, in the thermodynamic limit, reproduces Eq.~\eqref{Eq:E BG corr}.
The condition $y = \gamma N^2 \gg 1$ is equivalent to requiring that the healing length~\eqref{Eq:healing} be smaller than the size $L$ of the system.

The ground-state energy contains three contributions: 
the leading term corresponds to the usual mean field energy, 
the second contribution arises from the quantum fluctuations and is a one-dimensional analog of the  Lee-Huang-Yang correction in 3D, 
while the last term accounts for finite-size effects and depends explicitly on the interaction parameter $\gamma$.

\begin{figure}[htbp]
\centering
\includegraphics[scale=0.5]{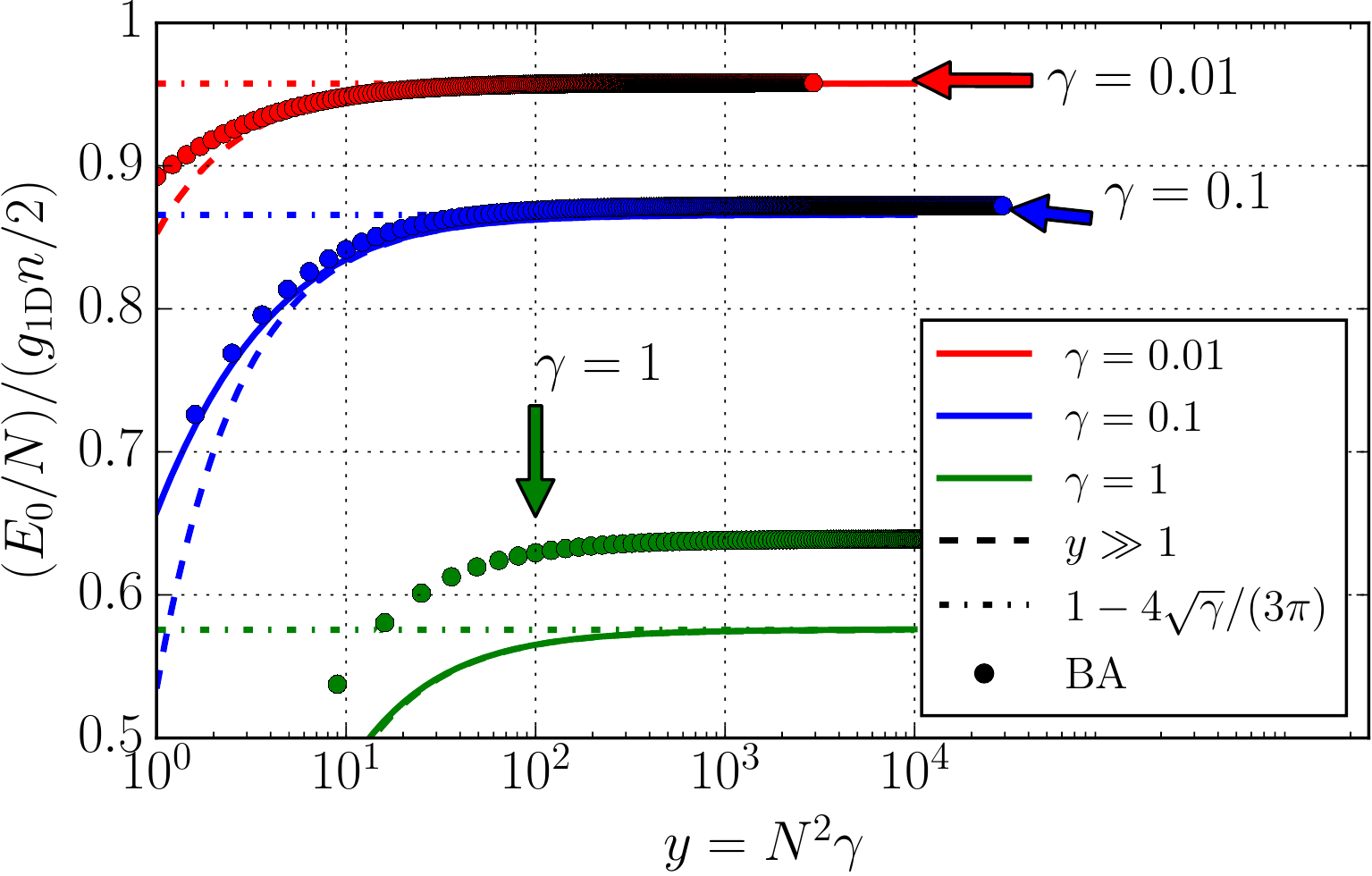}
\caption{(Color online) Comparison of the ground state energy per particle, in BG units, as a function of $y= \gamma N^2$ in the thermodynamic limit of Bogoliubov theory~\eqref{Eq:E BG corr} (dot-dashed line), the Bethe-{\it ansatz} (BA) calculation (circle), the Bogoliubov expression~\eqref{Eq:E G} (solid line) and the $y \gg 1$ expansion~\eqref{Eq:E BG ring limit} (dashed line), for several values of the interaction parameter $\gamma$.\label{fig:EBGtot}}
\end{figure}

Finite size corrections can be sizeable, as clearly shown by Fig.~\ref{fig:EBGtot} where we report the energy per particle as a function of $y$ for the thermodynamic limit~\eqref{Eq:E BG corr} (dot-dashed line), the Bethe-{\it ansatz} (BA) calculation (circle), the Bogoliubov expression~\eqref{Eq:E G} (solid line) and the expansion~\eqref{Eq:E BG ring limit} (dashed line).
The figure reveals a general good agreement between the BA and the Bogoliubov predictions~\eqref{Eq:E G}, except for $\gamma = 1$, where Eq.~\eqref{Eq:E G}, being based on the Bogoliubov approach, is no longer adequate.

\begin{figure}[htbp]
\centering
\includegraphics[scale=0.5]{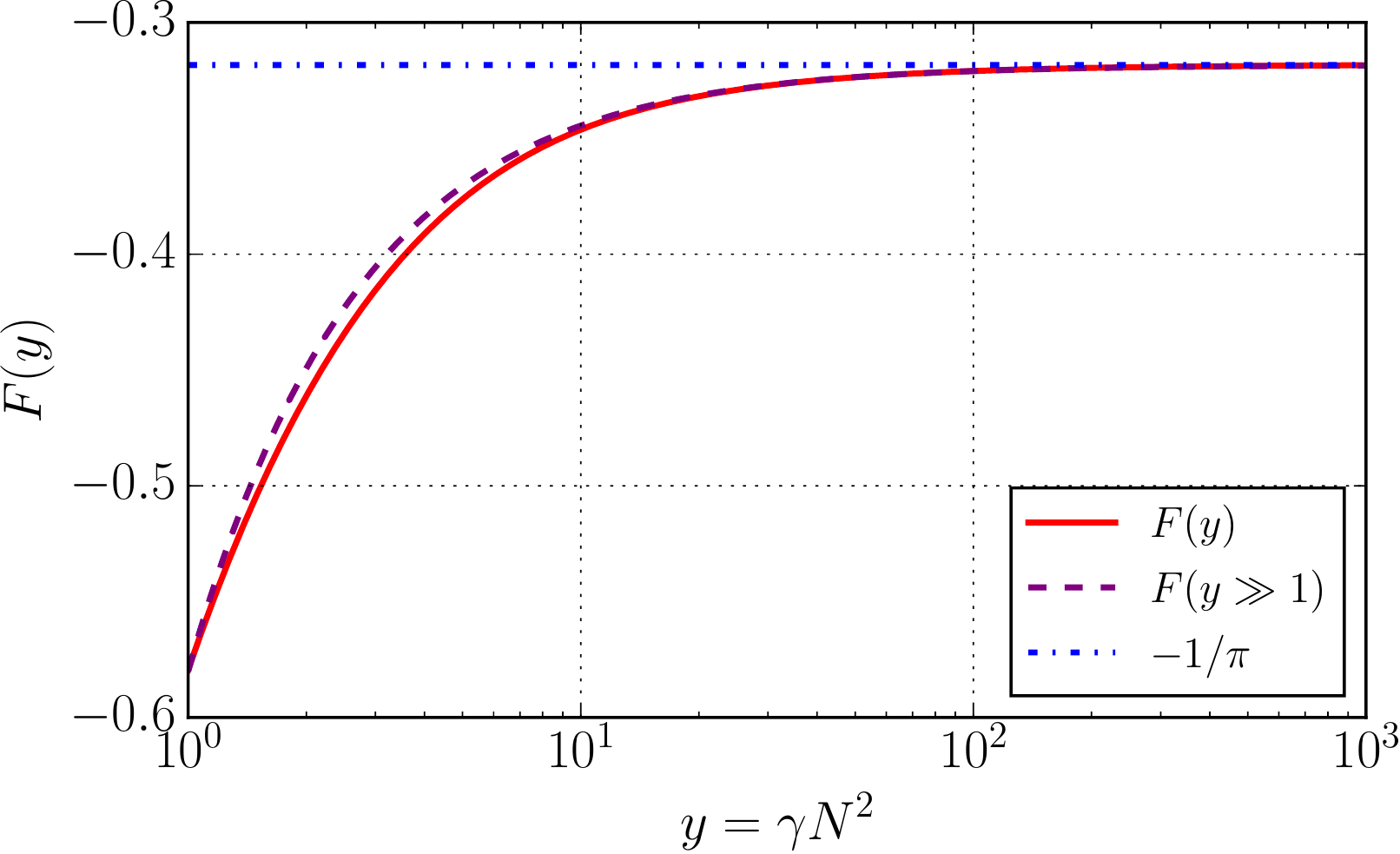}
\caption{(Color online) Comparison of the numerical series $F(y)$~\eqref{Eq:F} (solid line) and its analytical expansion~\eqref{Eq:FL} (dashed line) holding for $y \gg 1$. The dot-dashed line represents the thermodynamic value.\label{fig:F}}
\end{figure}

The chemical potential can be obtained by deriving Eq.~\eqref{Eq:E no self} with respect to $N$, at fixed $L$. One finds
\begin{equation}
\label{Eq:mu no self}
\mu = \left(\frac{\partial E_0}{\partial N}\right)_L = g_{\rm 1D}n \left[1 + \frac{1}{2N} \sum_{p = -\infty}^{+\infty} \left(\frac{p^2}{2m}\frac{1}{\epsilon(p)} - 1  \right)\right]
\end{equation}
which can be rewritten as $\mu =g_{\rm 1D}n  [1+ \sqrt{\gamma} F(y)]$, where $y$ is provided by Eq.~\eqref{Eq:y} and we have introduced the series
\begin{equation}
\label{Eq:F}
F(y) =\frac{1}{\sqrt{y}}\sum_{n_i = 0}^{+\infty} \left( \frac{ \pi n_i}{\sqrt{y +(n_i \pi)^2}}-1\right) + \frac{1}{2\sqrt{y}}
\end{equation}
depending on the quantized momenta~\eqref{Eq:quantized p} and such that the zero-momentum term is accounted for once.
The Euler-Maclaurin expression, applied to the sum~\eqref{Eq:F}, yields
\begin{equation}
\label{Eq:FL}
F(y \gg 1) \approx - \frac{1}{\pi} -  \frac{\pi}{12y}
\end{equation}
holding in the $y \gg 1$ limit.
In Fig.~\ref{fig:F} we report the comparison of the series~\eqref{Eq:F} with its expansion~\eqref{Eq:FL} holding for $y \gg 1$. The two curves agree very well for $y > 10$.
 
Using Eq.~\eqref{Eq:FL}, one can finally write the following expansion for the chemical potential
\begin{equation}
\label{Eq:muL}
\mu(\gamma N^2 \gg 1, \gamma \ll 1) \approx g_{\rm 1D}n \left[ 1 -  \frac{\sqrt{\gamma}}{\pi}  - \frac{\pi}{12N^2 \sqrt{\gamma}}\right] \ .
\end{equation}

\begin{figure}[htbp]
\centering
\includegraphics[scale=0.5]{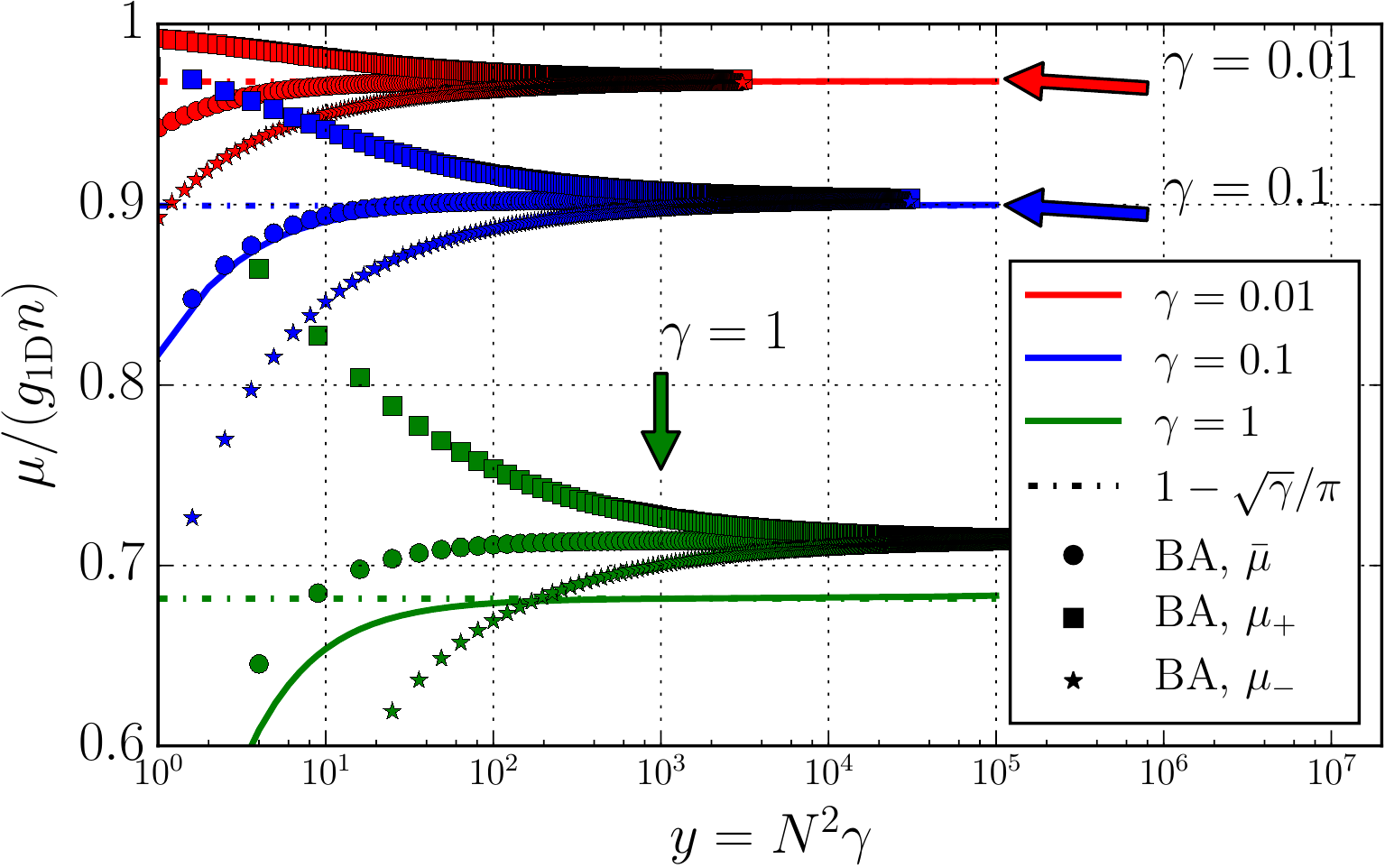}
\caption{(Color online) Comparison of the chemical potential in BG units as a function of $y = \gamma N^2$ in the thermodynamic limit~\eqref{Eq:mu BG corr} (dot-dashed line), the Bethe-{\it ansatz} (BA) calculation (symbols) and the Bogoliubov expression~\eqref{Eq:mu no self} (solid line), for several values of the interaction parameter $\gamma$. For the BA: $\mu_+ = E_0(N + 1) - E_0(N) $ (square), $\mu_- = E_0(N) - E_0(N - 1) $ (star) and $\bar{\mu} = (\mu_+ + \mu_-)/2$ (circle). \label{fig:muBGtot}}
\end{figure}

In Fig.~\ref{fig:muBGtot} we report the results for the chemical potential as a function of $y$ \eqref{Eq:y} for the thermodynamic limit~\eqref{Eq:mu BG corr} (dot-dashed line), the Bethe-{\it ansatz} calculation (symbols) and the Bogoliubov expression~\eqref{Eq:mu no self} (solid line). The $y \gg 1$ expansion~\eqref{Eq:muL} practically coincides with the full series~\eqref{Eq:mu no self}. The square symbol corresponds to the ``forward'' definition $\mu_+ = E_0(N + 1) - E_0(N)$ of the chemical potential, the star symbol to the ``backward'' expression $\mu_- = E_0(N) - E_0(N - 1) $, while the circles to the ``symmetric'' value $\bar{\mu} = (\mu_+ + \mu_-)/2$. 
While the three definitions of the chemical potential coincide in the thermodynamic limit $N \to \infty$, they are different in a finite system ~\cite{nuclear}. In particular, the symmetric definition $\bar{\mu}$ well agrees with the calculation~\eqref{Eq:mu no self}, based on the differential definition $\mu = (\partial E_0/\partial N)_L$, except for the $\gamma = 1$ case. 

From Eq.~\eqref{Eq:mu no self}, one can also calculate the sound velocity~\eqref{Eq:vs}, corresponding to the density derivative of the chemical potential for a fixed value of $L$. 
The resulting expression,
\begin{equation}
\label{Eq:v no self}
v_s(\gamma) = v_s^{\rm BG}(\gamma)\sqrt{1 - \frac{ g_{\rm 1D}n}{2N}\sum_{p = - \infty}^{+\infty}\left(\frac{p^2}{2m}\right)^2\frac{1}{\epsilon^3(p)}}
\end{equation}
with $v_s^{\rm BG}(\gamma)$ the sound velocity defined in the Bogoliubov regime, 
used in Fig.~\ref{fig:v}.
The above expression can be rewritten as $v_s(\gamma) = v_s^{\rm BG}(\gamma)\sqrt{1 - \sqrt{\gamma}H(y)}$ where we have defined the series
\begin{equation}
\label{Eq:H}
H(y) = \frac{\sqrt{y}}{2}\sum_{n_i = 0}^{+\infty} \frac{\pi n_i}{\left[y + (\pi n_i)^2 \right]^{3/2}}\ ,
\end{equation}
after introducing the variable $y$~\eqref{Eq:y} and the quantized momenta~\eqref{Eq:quantized p}.
As before, we apply the Euler-Maclaurin formula and we find the expansion
\begin{equation}
\label{Eq:HL}
H(y \gg 1) \approx \frac{1}{2\pi} - \frac{\pi}{24 y}
\end{equation}
holding in the $y \gg 1$ limit, yielding the asymptotic expansion

\begin{equation}
\label{Eq:vL gamma small}
v_s(\gamma N^2 \gg 1, \gamma \ll 1) \approx v_{s}^{\rm BG}(\gamma)\sqrt{1 - \frac{\sqrt{\gamma}}{2\pi} + \frac{\pi}{24N^2 \sqrt{\gamma}}}
\end{equation}
for the sound velocity.

\subsection{Tonks--Girardeau regime at $T = 0$}

\begin{figure}[htbp]
\centering
\includegraphics[scale=0.56]{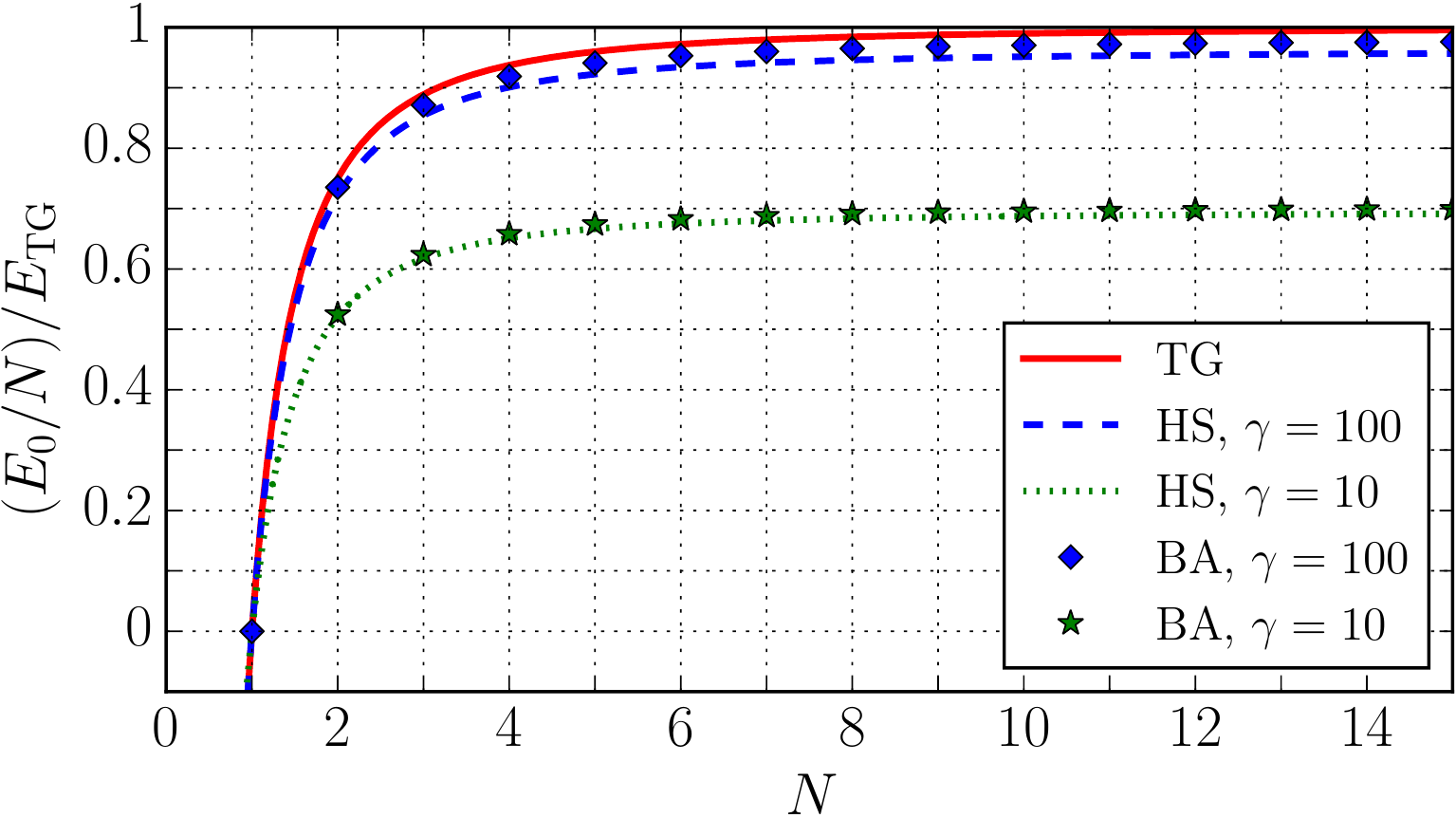}
\caption[Energy per particle as a function of $N$ for BA, TG and hard--spheres]{(Color online) Energy per particle in units of the TG gas energy $E_{\rm TG} = \pi^2 \hbar^2 n^2/(6m)$ as a function of $N$. Bethe--{\it ansatz} (BA) results (symbols) with different values of the interaction parameter $\gamma$ are compared with the  TG gas~\eqref{Eq:IFG energy N} (solid line) and the hard--sphere (HS) model~\eqref{Eq:E T0 N TG} (dashed and dotted lines). \label{fig:ETGN}}
\end{figure}

According to Girardeau~\cite{Girardeau1960}, the ground-state energy of the gas in the strongly-interacting limit is the same as that of an ideal Fermi gas. 
The energy for a finite number of particles $N$ in a box with periodic boundary conditions is obtained by summing the energy of the single-particle levels in the box,
\begin{equation}
\label{Eq:IFG energy N}
\frac{E_0}{N}(N) = \frac{\hbar^2}{mN}\sum_{n_i = 1}^{\frac{1}{2}(N - 1)}\left(\frac{2 \pi n_i}{L} \right)^2 = \frac{1}{6} \left(1 - \frac{1}{N^2}  \right)\frac{\pi^2 \hbar^2 n^2}{m}\ .
\end{equation}
In the thermodynamic limit, $N = \infty$, Eq.~(\ref{Eq:IFG energy N}) results in $E_{\rm TG} = \pi^2 \hbar^2 n^2/(6m)$. 
The ``excluded volume'' correction should be present for a finite interaction strength, see the hard--sphere like expression, Eq.~(\ref{Eq:E hard sphere}), and the discussion below it.
In order to incorporate the leading finite-size correction close to the Tonks-Girardeau regime we replace $L$ with $L-Na_{1D}$ in Eq.~\eqref{Eq:IFG energy N} resulting in the following expression for the energy per particle
\begin{equation}
\label{Eq:E T0 N TG}
\frac{E_0}{N}(N, \gamma) = \frac{1}{6}\frac{\pi^2 \hbar^2 n^2}{m} \left(1 - \frac{1}{N^2}  \right) \left(1 + \frac{2}{\gamma} \right)^{-2}\ .
\end{equation}
For large values of the interaction parameter $\gamma$ one can replace the factor $\left(  1 + 2/\gamma \right)^{-2}$ with $\left(1 - 4/\gamma \right)$.
In Fig.~\ref{fig:ETGN} we report the energy per particle as a function of $N$ for the TG regime~\eqref{Eq:IFG energy N} (solid line), the hard-sphere (HS) like model~\eqref{Eq:E T0 N TG} (dashed and dotted lines) and the Bethe-{\it ansatz} solution (symbols) for several values of $\gamma$. We observe a very good agreement between the BA solution and the analytical hard-sphere~\eqref{Eq:E T0 N TG} expression. For $\gamma = 1000$ the BA results are indistinguishable from the TG limit~\eqref{Eq:IFG energy N} and they are not reported in the figure. The comparison between Eq.~\eqref{Eq:E T0 N TG} and Eq.~\eqref{Eq:E BG ring limit} reveals that finite-size effects are less important in the TG regime since in the weakly interacting Bogoliubov regime, the correction $1/(N^2 \sqrt{\gamma})$ is amplified by the smallness of $\gamma$.

For strong repulsion we obtain the finite-size correction to the chemical potential
\begin{equation}
\label{Eq:mu T0 N TG}
\mu(N, |\gamma| \gg 1) = \left(\frac{\partial E_0}{\partial N}\right)_L \approx  E_F \left[1 - \frac{16}{3\gamma} - \frac{1}{3N^2} \left(1 - \frac{8}{\gamma}\right) \right]
\end{equation}
and to the sound velocity~\eqref{Eq:vs}:
\begin{equation}
\label{Eq:v T0 N TG}
v_s(N, |\gamma| \gg 1) \approx v_F \sqrt{1 - \frac{8}{\gamma} + \frac{4}{3\gamma N^2}}\ .
\end{equation}
It is interesting to note that while the finite-size correction to the energy~(\ref{Eq:E T0 N TG}) and the chemical potential~(\ref{Eq:mu T0 N TG}) scales as $1/N^2$ with the number of particles, such a correction is instead asymptotically vanishing in the sound velocity~\eqref{Eq:v T0 N TG}. 

\begin{figure}[htbp]
\centering
\includegraphics[scale=0.55]{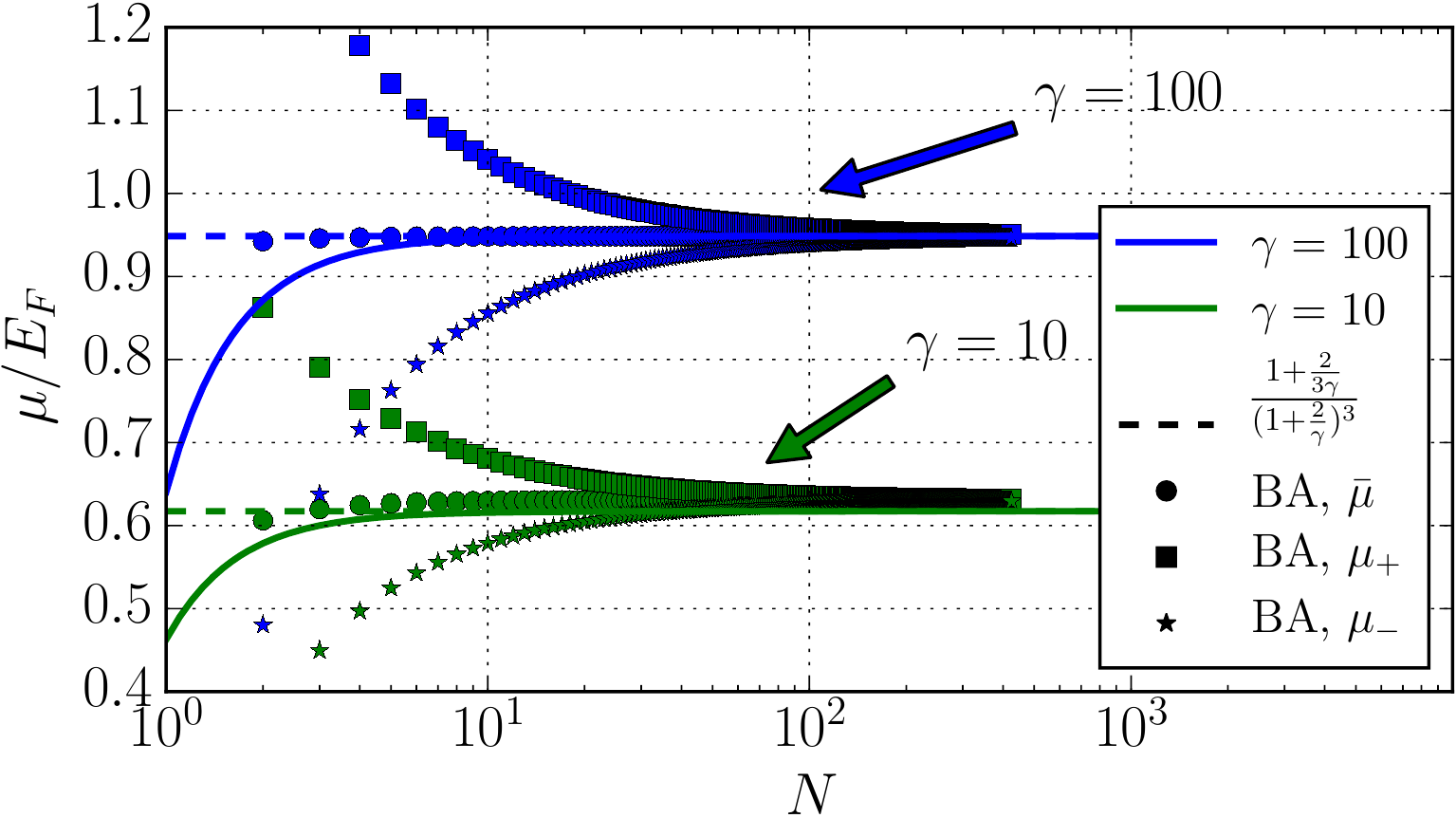}
\caption{(Color online) Chemical potential at $T = 0$ in Fermi units as a function of the number of particles $N$ in the TG regime \eqref{Eq:muexactnote} for fixed values of $\gamma$ (solid line). The dashed lines correspond to the thermodynamic limit $[1 + 2/(3\gamma)]/(1 + 2/\gamma)^3$ of the TG model~\eqref{Eq:TGthermonote}. The symbols correspond to the Bethe-{\it ansatz} (BA) calculation: $\mu_+ = E_0(N + 1) - E_0(N) $ (square), $\mu_- = E_0(N) - E_0(N - 1) $ (star) and $\bar{\mu} = (\mu_+ + \mu_-)/2$ (circle).\label{fig:alphaNTG}}
\end{figure}

In Fig.~\ref{fig:alphaNTG}, we plot the chemical potential $\mu = (\partial E_0/\partial N)_{|L}$ with $E_0$ given by Eq.~\eqref{Eq:E T0 N TG} (solid line) as a function of $N$ for different values of $\gamma$~\cite{muTGring}. In the same figure we plot also the values of $\mu_+$ and $\mu_-$ which differ from the symmetric value $\bar{\mu}=(\mu_++\mu_-)/2$ for small values of $N$ \cite{nuclear}, similarly to the case of the weakly interacting Bose gas. Differently from the weakly interacting BG gas, the symmetric value $\bar{\mu}$ however exhibits significant deviations with respect to the differential estimate $(\partial E_0/\partial N)_{|L}$, for small values of $N$.

\subsection{Static inelastic structure factor}

The ring geometry has a profound effect on the correlation functions.
Here we analyze the inelastic static structure factor at zero temperature,
\begin{equation}
S(k) = \frac{1}{N}\left[\langle \rho_{k}\rho_{-k}\rangle - |\langle \rho_{k}\rangle|^2\right]
\label{Eq:Sk}
\end{equation}
where $\rho_k = \sum_{j=1}^N e^{-ikx_j}$ is the density operator in momentum representation.
The static structure factor gives information about two-body correlations and can be measured in experiments by means of Bragg spectroscopy. 

In the thermodynamic limit, the static structure factor has a linear behavior at small momenta, $S(k) = \hbar |k| / (2mv_s)$, with the slope determined by the sound velocity $v_s$.
The ring geometry introduces both discretization in the allowed momentum and a change in the slope due to the finite-size correction to the sound velocity.
The latter effect is rather weak, especially in the Tonks-Girardeau regime, but is important in the context of the finite-size dependence of the Luttinger parameter $K_L(N)$.

 The strongest effect comes from the discretization of the allowed momenta on a ring.
For the standing wave values~\eqref{Eq:quantized p}, the last term in Eq.~(\ref{Eq:Sk}), corresponding to the square of the so-called elastic form factor, does not contribute.
Indeed, one finds
$|\langle \rho_{k}\rangle|^2 / N  = N |\langle e^{ikx}\rangle|^2 = N[\sin(kL/2) / (kL/2)]^2$, which exactly vanishes for $k = n_i$. 

\begin{figure}[htbp]
\centering
\includegraphics[width=\columnwidth]{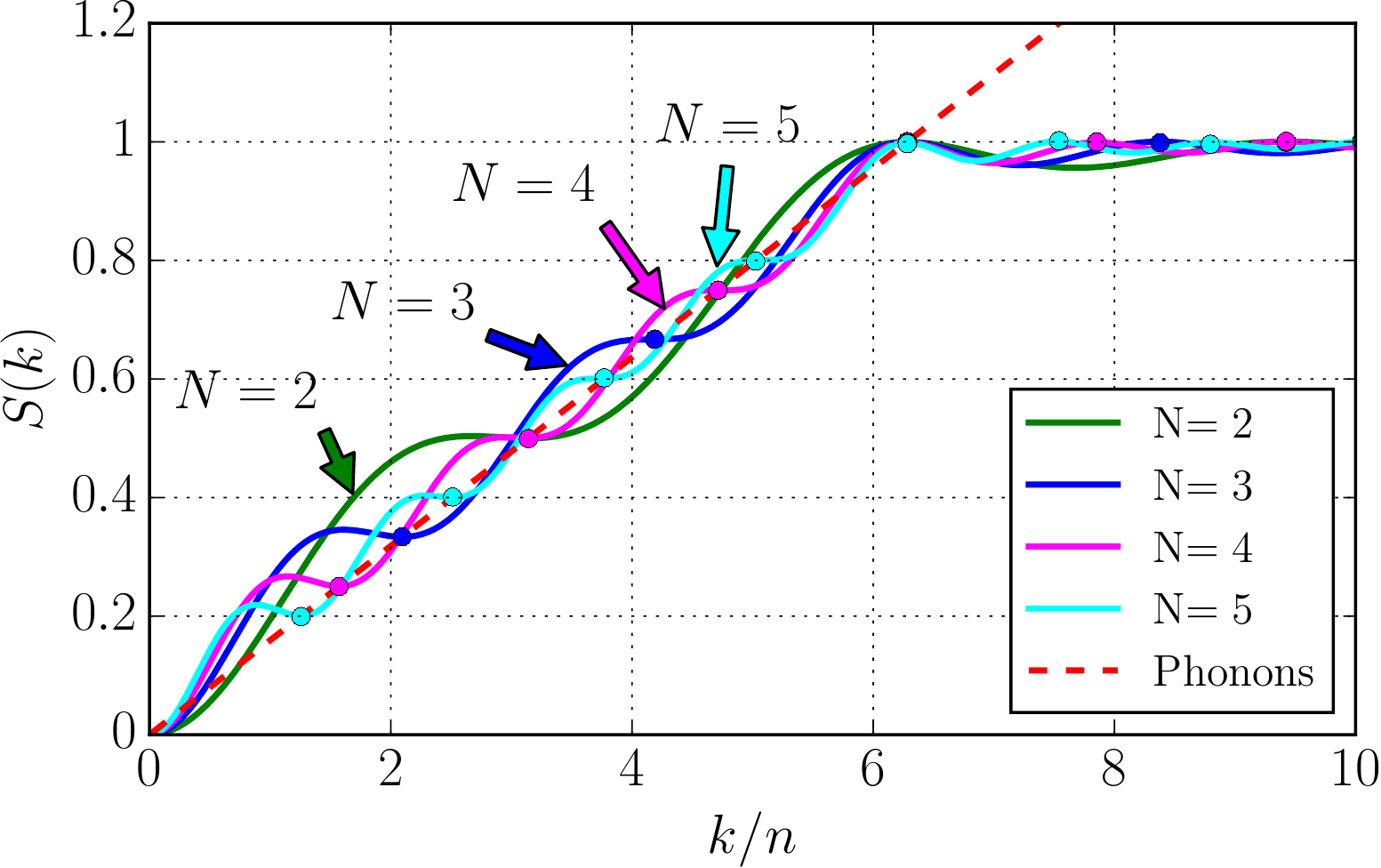}
\caption{(Color online) Static structure factor at $T=0$ in the Tonks-Girardeau limit for different number of particles (solid lines).
Values of momenta~\eqref{Eq:quantized p} $k_i = 2\pi n_i/L$ corresponding to standing waves on a ring are shown with circles.
The dashed line represents the phononic law $S(k) = \hbar |k| /(2 m v_F)$, which coincides with the static structure factor in the thermodynamic
limit for $|k|<2\pi n$ in the TG regime.
\label{fig:Sk:TG}}
\end{figure}

Figure~\ref{fig:Sk:TG} reports the static structure factor in the Tonks-Girardeau regime.
When the probing momentum $k$ is equal to a standing wave value~\eqref{Eq:quantized p} in the ring, the value of the static structure factor is exactly the same as in the thermodynamic limit.
In this way, discrete $S(k_i)$ points form a linear phononic dependence.
As the number of particles is increased, the phononic behavior is better resolved.
The absence of the change of the slope means that the finite-size corrections to the sound velocity are negligible in the Tonks-Girardeau regime, confirming the predictions of Eq.~(\ref{Eq:v T0 N TG}).

When the probing momentum $k$ is different from the allowed values in the ring, the value of $S(k)$ depends strongly on the number of particles.
Importantly, the small-momentum behavior is no longer linear but rather shows a quadratic dependence on $k$.
This qualitative change reflects the change in the structure of the excitation spectrum which becomes discrete.
A quadratic dependence on the momentum, $S(k) = \hbar^2k^2/(2m\Delta)$, is typical to gapped systems with $\Delta$ being the value of the gap.
In the discrete case it is not possible to create an excitation with energy smaller than $\Delta \propto \hbar^2/(mL^2)$, resulting in a quadratic low-momentum dependence.
In the thermodynamic limit $\Delta\to 0$ and the phononic linear behavior is restored.

\begin{figure}[htbp]
\centering
\includegraphics[width=\columnwidth]{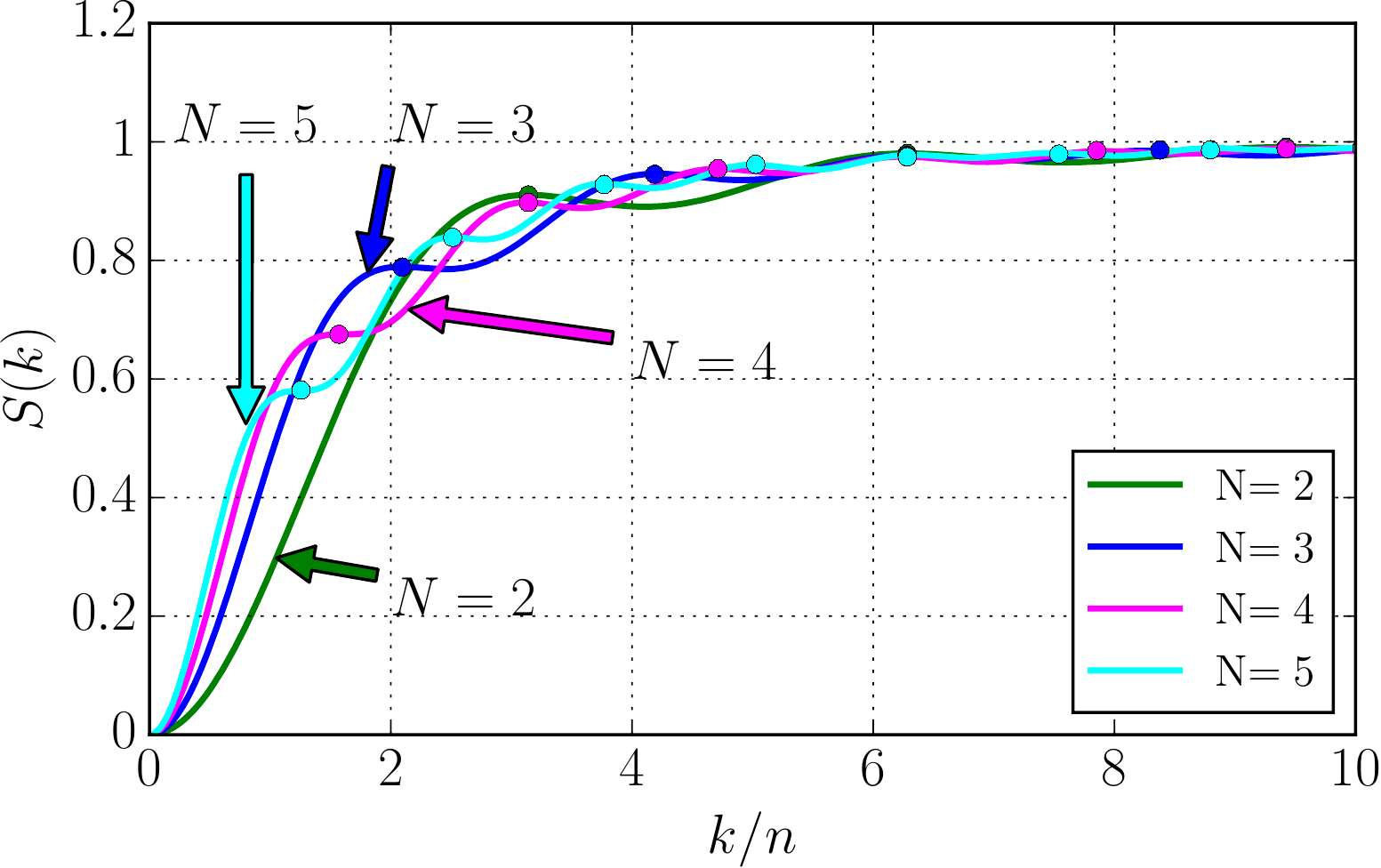}
\caption{(Color online) Static structure factor at $T=0$ and $\gamma = 1$ for different number of particles (solid lines).
Values of momenta~\eqref{Eq:quantized p} $k_i = 2\pi n_i/L$ corresponding to standing waves on a ring are shown with circles.
\label{fig:Sk:gamma1}}
\end{figure}

In Fig.~\ref{fig:Sk:gamma1} we show the static structure factor for $\gamma = 1$, calculated using the diffusion Monte Carlo method.
Similarly, the finite-size quadratic behavior at small momenta is replaced by the linear phononic dependence in the thermodynamic limit.
Contrarily to the TG case, here the values at $S(k_i)$ depend on the number of particles, although the effect is weak (see, for example, the value at $k=\pi n$).
In terms of the Luttinger parameter, which in the linear regime corresponds to $K_L = 2\pi n S(k) / k$, this results in its finite-size dependence.

While for the TG regime, the linear dependence extends up to $k=2\pi n$, for weaker interactions the linear regime shrinks (compare Figs.~\ref{fig:Sk:TG}-\ref{fig:Sk:gamma1}).
Eventually for $\gamma\to 0$ the linear regime becomes very small and phononic theory cannot provide a good description of the system properties.
A similar effect was observed in Figs.~\ref{fig:comparison}-\ref{fig:mu_gn} in the applicability of the phononic theory in the limit of weak interactions.

\section{Conclusion}
\label{Sec:Conclusion}

In this paper we have investigated the low temperature properties of 1D Bose gases along the whole Bogoliubov (BG) --- Tonks-Girardeau (TG) crossover. We have shown that, at low temperature, the chemical potential exhibits a typical $T^2$ behavior, which follows from the leading contribution to thermodynamics arising from the thermal excitation of phonons, similarly to what happens in superfluids. 
The chemical potential is always a decreasing function of $T$ at high
temperature, thus the $T^2$ increase exhibited by the chemical
potential at low temperature is responsible for a typical
non-monotonic behavior as a function of $T$.
The coefficient of the $T^2$ law has been calculated using the Lieb-Liniger results for the sound velocity and the resulting behavior has been successfully compared with thermodynamic functions obtained from the Yang-Yang theory of 1D interacting Bose gases. We have also presented results for the temperature dependence of the isothermal and adiabatic inverse compressibilities. In particular we have shown that the $T^2$ correction has opposite sign in the two cases.

In the second part of the paper we have focused on the corrections to the thermodynamic functions caused by the finite size of the system. To this purpose, we have considered the useful ring geometry and the mapping with the 1D problem where calculations are carried out using periodic boundary conditions. Explicit results have been obtained in the weakly and strongly interacting regimes where, at zero temperature, the first corrections to the thermodynamic limit, due to finite size effects, can be calculated in analytic form, in excellent agreement with the numerical results provided by the Bethe-{\it ansatz}.  We have found that finite-size corrections are particularly important in the weakly interacting regime where the healing length can easily  become comparable to the size of the system.

Concerning future developments of the analysis carried out in this paper, it is worth mentioning the physical understanding of higher-order corrections (beyond the $T^2$-law caused by the real excitations of the phononic branch) to the low-temperature thermodynamic behavior. 
In particular, it is important to understand the temperature
corrections arising due to non-symmetric spreading of the phononic
branch (different beyond-linear behavior of the lower and upper
branches) as well as effects originating from the non-linear behavior of
the Bogoliubov spectrum at large momenta. 
A further perspective of research concerns the finite temperature thermodynamic behavior of 1D Bose gases containing a small number of atoms and confined in a ring of finite size.

\appendix

\section{Euler-Maclaurin expansion for G(y)}
\label{Sec:G}

In this Appendix, we show the detailed derivation of the expansion holding for $y \gg 1$~\eqref{Eq:GL} for the series~\eqref{Eq:G}.

We use the Euler-Maclaurin expansion which allows to approximate a series as follows~\cite{Abramowitz2012}:
\begin{equation}
\label{Eq:S-I G}
\sum_{k = 0}^{+ \infty} f(k) \approx \int_0^{+\infty} f(x) dx + \sum_{k = 1}^{m}\frac{B_{k}}{k!}f^{(k - 1)}(x)|_{0}^{+\infty} 
\end{equation}
where $f(x)$ is a continuous function of real numbers $x$ in the interval $[0, + \infty]$. For $m = 2$, one considers only the first terms in the sum, whose Bernoulli's numbers are
\begin{equation}
\label{Eq:Bernoulli G}
\begin{cases}
B_1 = -\frac{1}{2} \\
B_2 = \frac{1}{6} \\
\end{cases}
\end{equation}
and $f^{(k)}(x)$ are the $k$--derivatives of the function $f(x)$.

By defining the following function
\begin{equation}
\label{Eq:f G}
f(x) = 2 \pi x \sqrt{y + (\pi x)^2} - 2(\pi x)^2 - y
\end{equation}
entering the series~\eqref{Eq:G}, one estimates the integral
\begin{equation}
\label{Eq:I G}
\int_0^{+\infty} dx f(x) = - \frac{2 y \sqrt{y}}{3 \pi}\ ,
\end{equation}
which allows to calculate the thermodynamic limit of the ground-state energy per particle on a ring configuration~\eqref{Eq:E G}, provided by Eq.~\eqref{Eq:E BG corr}.

By calculating the first derivative of the function~\eqref{Eq:f G}, and by using Eq.~\eqref{Eq:S-I G} and Eq.~\eqref{Eq:Bernoulli G}, one finally gets the expansion~\eqref{Eq:GL} holding for large values of the $y$ parameter.

\begin{acknowledgments}
G. De Rosi and S. Stringari would like to acknowledge fruitful and helpful discussions with L. P. Pitaevskii, C. Menotti, S. Giorgini, M. Di Liberto and G. Bertaina. This work has been supported by ERC through the QGBE grant, by the QUIC grant of the Horizon2020 FET program and by Provincia Autonoma di Trento (De Rosi \& Stringari).

G. De Rosi acknowledges the hospitality of the Computer Simulation in Condensed Matter Research Group (SIMCON) of Universitat Polit\`{e}cnica de Catalunya in Barcelona, where this work was partially done.

G. E. Astrakharchik acknowledges partial financial support from the MICINN (Spain) Grant No.~FIS2014-56257-C2-1-P. The Barcelona Supercomputing Center (The Spanish National Supercomputing Center -- Centro Nacional de Supercomputaci\'on) is acknowledged for the provided computational facilities.

The authors gratefully acknowledge the Gauss Centre for Supercomputing e.V. (\href{http://www.gauss-centre.eu/gauss-centre/EN/Home/home_node.html}{www.gauss-centre.eu}) for funding this project by providing computing time on the GCS Supercomputer SuperMUC at Leibniz Supercomputing Centre (LRZ, \href{https://www.lrz.de/}{www.lrz.de}).

The authors would like to acknowledge also G. Lang, X.-W. Guan and the referees of this manuscript for useful comments and suggestions which have allowed some improvements in the revised version of this work.

\end{acknowledgments}


\begin{thebibliography}{99}
\bibitem{muTGring}
The explicit expression for the chemical potential is
\begin{equation}
\label{Eq:muexactnote}
\mu(N, \gamma) = \frac{E_F}{\left(1 + \frac{2}{\gamma} \right)^3} \left[1 + \frac{2}{3\gamma} - \frac{1}{3N^2}\left(1 - \frac{2}{\gamma} \right) \right]
\end{equation}
which, in the thermodynamic limit, gives
\begin{equation}
\label{Eq:TGthermonote}
\mu(\gamma) = \frac{E_F}{\left(1 + \frac{2}{\gamma} \right)^3} \left(1 + \frac{2}{3\gamma}\right)
\end{equation}
and for large values of the interaction parameter $\gamma$ provides Eq.~\eqref{Eq:mu T0 N TG}.
\end{thebibliography}
\end{document}